\DocumentMetadata{}
\documentclass[twocolumn]{aastex701}
\ExplSyntaxOn
\AddToHook{env/sidewaystable*/end}
  {\pdfmanagement_add:nnn{ThisPage}{Rotate}{90}}
\ExplSyntaxOff

\newcommand{\neii}{[Ne\,{\scriptsize II}]}
\newcommand{\neiii}{[Ne\,{\scriptsize III}]}
\newcommand{\arii}{[Ar\,{\scriptsize II}]}
\newcommand{\hi}{H\,{\scriptsize I}}
\newcommand{\feii}{[Fe\,{\scriptsize II}]}

\usepackage{caption}
\usepackage{comment}
\usepackage{array}

\begin{document}

\title{Discovery of Externally Irradiated ``Molecular Proplyds'' in the Low-FUV Upper Scorpius star-forming region}

\author[orcid=0000-0003-3401-1704,sname='Bajaj']{Naman S. Bajaj}
\affiliation{Lunar and Planetary Laboratory, The University of Arizona, Tucson, AZ 85721, USA}
\email[show]{namanbajaj@arizona.edu}  
\correspondingauthor{Naman S. Bajaj}

\author[0000-0003-0330-1506]{Andrew D. Sellek}
\affiliation{Université Paris-Saclay, CNRS, Institut d’Astrophysique Spatiale, 91405 Orsay, France}
\email[]{andrew-david.sellek@universite-paris-saclay.fr}

\author[0009-0004-8091-5055]{Rossella Anania}
\affiliation{School of Physics, Trinity College Dublin, The University of Dublin, College Green, Dublin 2, Ireland}
\email[]{ananiar@tcd.ie}

\author[0000-0001-7962-1683]{Ilaria Pascucci}
\affiliation{Lunar and Planetary Laboratory, The University of Arizona, Tucson, AZ 85721, USA}
\affiliation{Steward Observatory, The University of Arizona, 933 N. Cherry Avenue, Tucson, AZ 85721-0065, USA}
\email[]{pascucci@arizona.edu}

\author[0000-0001-6072-9344]{Jinyoung Serena Kim}
\affiliation{Steward Observatory, The University of Arizona, 933 N. Cherry Avenue, Tucson, AZ 85721-0065, USA}
\email[]{serena00@arizona.edu}

\author[0000-0002-7154-6065]{Gregory J. Herczeg}
\affiliation{Kavli Institute for Astronomy and Astrophysics, Peking University, Beijing 100871, People’s Republic of China}
\affiliation{Department of Astronomy, Peking University, Beijing 100871, People’s Republic of China}
\email[]{gherczeg1@gmail.com}

\author[0000-0001-6410-2899]{Richard Alexander}
\affiliation{School of Physics and Astronomy, University of Leicester, University Rd, Leicester, LE1 7RH, UK}
\email[]{richard.alexander@leicester.ac.uk}  

\author[0000-0001-7255-3251]{Gabriele Cugno}
\affiliation{Department of Astrophysics, University of Zürich, Winterthurerstrasse 190, 8057 Zürich, Switzerland}
\email[]{gabriele.cugno@uzh.ch}

\author[0000-0002-4022-4899]{Sierra L. Grant}
\affiliation{Earth and Planets Laboratory, Carnegie Institution for Science, 5241 Broad Branch Road, NW, Washington, DC 20015, USA}
\email[]{sgrant@carnegiescience.edu}

\author[0009-0007-0600-6461]{Sophie C. Clark}
\affiliation{Lunar and Planetary Laboratory, The University of Arizona, Tucson, AZ 85721, USA}
\email[]{sophieclark@arizona.edu}

\begin{abstract}

External photoevaporation driven by UV radiation from massive stars is expected to affect a large fraction of protoplanetary disks, yet direct evidence has largely been confined to strongly irradiated environments ($>1000\,G_0$).
In nearby low-mass star-forming regions with weak UV fields ($<100\,G_0$), where much of our understanding of planet formation has been established, external photoevaporation has only been inferred indirectly. We report the first direct detection of externally photoevaporating disks in this regime.
We analyze archival JWST MIRI/IFU observations of 33 protoplanetary disks in the Upper Scorpius star-forming region and identify cometary H$_2$ emission morphology around three systems, tracing H$_2$ dissociation fronts tens to hundreds of au from the disks.
The fronts point directly toward nearby early B-type stars, and the measured stellar proper motions rule out a bow-shock origin, establishing these systems as the weakest-FUV-proplyds identified to date.
Two of the three host stars have masses $\leq0.15\,M_\odot$, suggesting enhanced susceptibility of disks around very low-mass stars to external photoevaporation.
From spatially resolved H$_2$ rotational-diagram fits, we estimate wind mass-loss rates of $\sim10^{-9}\,M_\odot\,\mathrm{yr}^{-1}$.
The \textsc{FRIEDv2} theoretical grid predicts comparable rates for several additional disks without detected fronts.
We propose that 3D disk--star geometry may account for this difference: direct illumination of the weakly bound outer edge could drive mass loss more efficiently than surface illumination in weak FUV fields.
These detections extend direct evidence for external photoevaporation to fields below $100\,G_0$ and establish spatially extended H$_2$ emission as a powerful new diagnostic of this process in nearby star-forming regions.

\end{abstract}


\section{Introduction} \label{sec:intro}

How protoplanetary disks evolve and disperse directly affects the final architecture of planetary systems. A vast majority of disks are affected by outside-in ``external photoevaporation'' (PE) caused by intense UV radiation from nearby massive (O- and B-type) stars leading to loss of gas from the disk \citep[e.g.,][]{Odell1998,Smith2005,Eisner2008,Robberto2013,Aru2024,Boyden2024}, with evidence in the form of smaller disk sizes \citep{Boyden2020,Otter2021,Ballering2023} and reduced disk masses \citep{Mann2014,Ansdell2017,vanterwisga2020,Mauco2023,vanterwisga2023}. This phenomenon was first observed in the heavily irradiated Orion Nebula Cluster \citep[ONC,][]{Mccaughrean1996,Johnstone1998,Odell1998,Bally2000,Odell2001,Henney2002} in the form of proplyds, i.e., disks surrounded by ionized gas with a cometary morphology forming the ionization front (I-front). The typical far ultraviolet (FUV) fields in the ONC are of the order $\sim$10$^5$~G$_0$ \citep[where 1~G$_0$ is the average interstellar value,][]{Habing1968}.

In the last decade, proplyds have been directly resolved in weaker fields down to $\sim$1000~G$_0$ \citep[e.g., around B-type stars,][]{Kim2016, Boyden2024,Booth2026}. A major open question is the extent to which external PE affects disks in nearby low-mass star-forming regions, where most disks are exposed to FUV radiation fields of $\leq$100~G$_0$ \citep[e.g.,][]{Winter2022,Anania2025a}. Studies of disk evolution and planet formation have so far focused on disks in these low-FUV regions; hence, understanding the impact of external PE here is crucial to gain a more complete picture. 

Theoretical studies have long predicted that external PE can influence disks in such environments at rates comparable to those of internal disk-evolution processes \citep[e.g.,][]{Facchini2016,Haworth2018,Sellek2020}, with one study finding potential evidence of external PE towards the very weakly irradiated IM~Lup \citep[4~G$_0$,][]{Haworth2017}. More recently, population-level disk evolution models of the Upper Scorpius star-forming region, where radiation fields span $\sim$1-100~G$_0$, have found evidence for disk truncation by external PE \citep{Anania2025b,Ping2026}. In particular, these studies found that viscous evolution alone could not readily reproduce the observed Upper Sco disk sizes. At the same time, \cite{Bajaj2026} reported a serendipitous discovery of two proplyd candidates in the same region based on the morphology of their extended H$_2$ emission.

In this letter, we present a detailed analysis of those two previously identified candidate proplyds to determine the origin of their extended emission. We also investigate a third candidate identified through an archival analysis of MIRI observations of disks in Upper Sco. Section~\ref{sec:sample_details_calibration} describes the sample, observations, and data calibration. In Section~\ref{sec:results}, we analyze H$_2$ intensity maps of all Upper Sco disks in the sample to identify candidate proplyds by means of bright H$_2$ emission fronts. For the three candidates, we examine the orientation of each bright front relative to nearby A- and B-type stars to confirm them as dissociation fronts in an externally-driven photoevaporative wind and estimate the observed H$_2$ mass-loss rates. We then compare these estimates with theoretical predictions, where possible. In Section~\ref{sec:discussion}, we discuss the implications of our findings for external PE in weak radiation fields and for the broader Upper Sco disk population. We summarize our conclusions in Section~\ref{sec:summary_conclusion}. The H$_2$ S(1), S(3), and S(5) line-intensity maps for all sources are provided in Section~\ref{sec:h2_line_maps}.

\section{Sample details and Calibration} \label{sec:sample_details_calibration}

Motivated by the discovery of two proplyd candidates in Upper Sco by \cite{Bajaj2026}, we extend the analysis to all disks in the Upper Scorpius star-forming region observed with the James Webb Space Telescope (JWST) Mid-Infrared Instrument (MIRI) in Integral Field Unit (IFU) mode during Cycles 1 through 3. Our primary sample comprises 28 disks, of which 15 have disk inclinations $> 40^\circ$ and were previously analyzed in \cite{Bajaj2026}. This primary sample covers stellar masses ranging from 0.1 to 1.5 M$_{\odot}$, the majority of which (24/28) are below 1 M$_{\odot}$. These disks were observed across four different programs: GO 1584 \citep{Salyk2021} and GTO 1282 \citep{Henning2017}, which together account for four targets, and GO 2970 \citep{Pascucci2023prop} and GO 3034 \citep{Zhang2023}, which contribute the majority of the sample (24/28). In addition to the primary sample, we analyze 5 very-low-mass/brown-dwarf disks with stellar masses $<$0.1~M$_{\odot}$ \citep{Grant2023} to complete the existing archival dataset, leading us to a total of 33 disks.

All disks were observed using four dithers, with exposure times designed to achieve a signal-to-noise ratio (SNR) $\gtrsim$ 100 at $\sim$16~\micron{}, enabling the detection of molecular lines from several volatile species. Observations were carried out in all three MIRI-MRS bands (SHORT, MEDIUM, LONG), and a dedicated background observation was not obtained for most sources. Additional observational details are provided in \citet[][]{Henning2024,Arulanantham2025,Raul2026,Xie2026}; Grant, S. et al. (in prep).

All observations were calibrated locally using the standard JWST pipeline. We downloaded the raw \texttt{uncal} files from the Barbara A. Mikulski Archive for Space Telescopes (MAST) and processed them with version~2.0.1 of the JWST calibration pipeline \citep{Bushouse2026}, the most recent release available as of July 06, 2026. Reference files were selected using CRDS version~13.1.16 with the context \texttt{jwst\_1535.pmap}. To ensure consistency across the sample, the same pipeline version and CRDS configuration were applied to all observations. Additional details on the calibration procedure can be found in \cite{Bajaj2026}.

\section{Results} \label{sec:results}

The goal of this study is to identify spatially resolved signatures of externally driven photoevaporative winds. Accordingly, in this section, we first present the analysis of atomic and molecular line intensity maps (Section~\ref{sec:atomic_molecular_intensity_maps}). Of the 33 disks analyzed, three show promising signatures of ongoing external PE \citep[two reported in][]{Bajaj2026}. For these three disks, we perform further analyses to confirm their externally irradiated nature (Section~\ref{sec:confirmation_proplyds}) and calculate both the observed and theoretically predicted wind mass loss rates (Sections~\ref{sec:h2_mass_loss_calculation} and \ref{sec:comparison_theoretical_pred_allsources}).

\subsection{Atomic and Molecular line intensity maps} \label{sec:atomic_molecular_intensity_maps}

\begin{figure*}
\centering
\includegraphics[width=\linewidth]{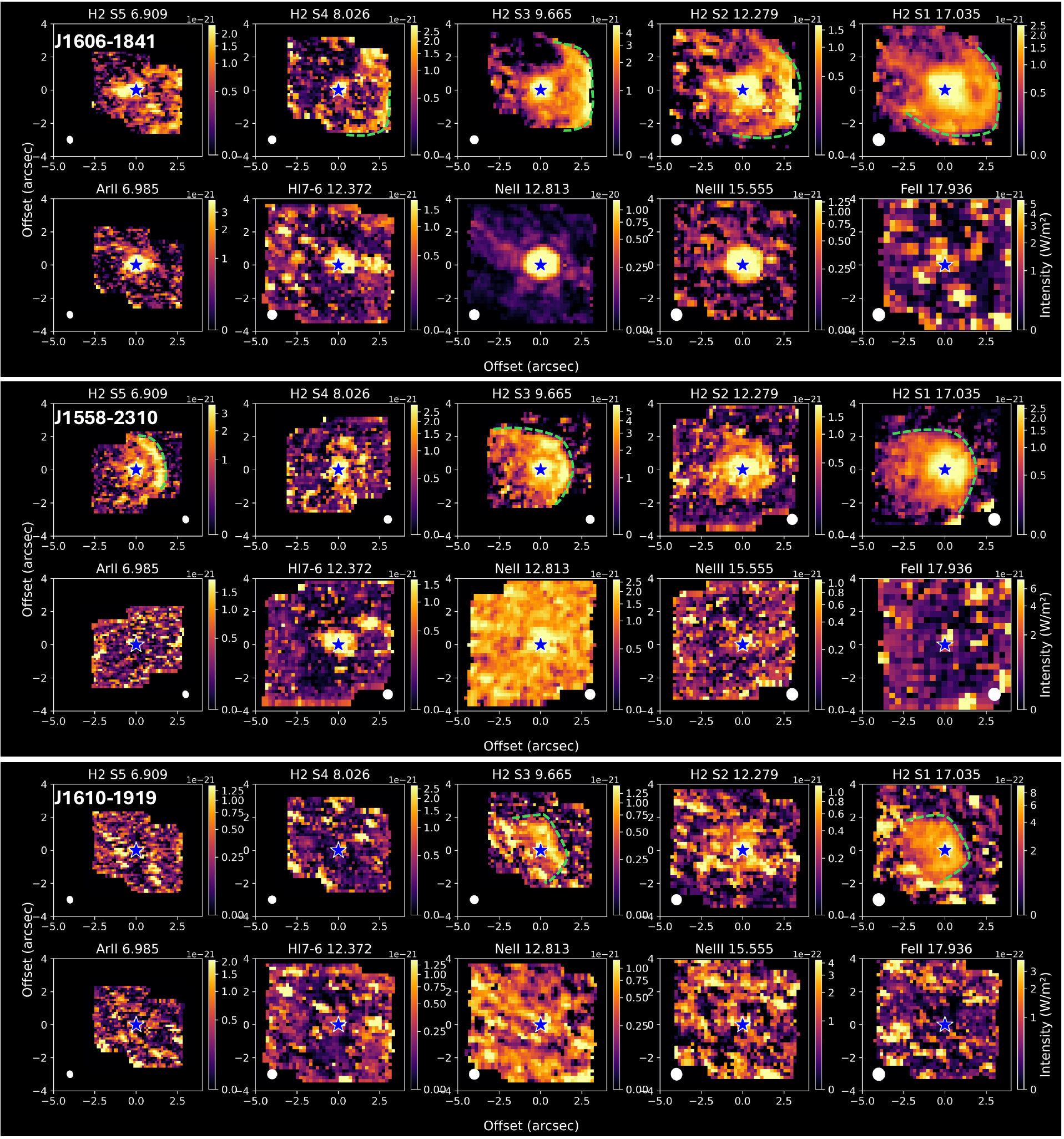}
\caption{Spaxel-by-spaxel continuum-subtracted molecular and atomic line intensity maps toward J1606-1841 (top), J1558-2310 (middle), and J1610-1919 (bottom), the three objects that show proplyd-like morphology in H$_2$. The blue star in each panel marks the corresponding continuum centroid (i.e., the stellar location). The white ellipse in the bottom corner represents the PSF FWHM at the line wavelength. Here, an arcsec approximately corresponds to $\sim$140-150~au depending on the distance to the source (Table \ref{tab:properties}), and the PSF FWHM can be estimated as 0.033 ($\lambda$/micron) + 0.''106 \citep{Law2023}. The green dashed curve is overlaid to guide the eye to the location of the dissociation front. Upper limits at 1 sigma are shown where the amplitude of the fitted Gaussian falls below 1$\sigma$, with $\sigma$ defined as the standard deviation of the flux in wavelength channels immediately outside the line. All the maps are plotted at their original spatial resolution.}
\label{fig:grid_maps}
\end{figure*}

\begin{deluxetable*}{c|cc|cc|cc}
\tablecaption{Literature source properties and MIRI H$_2$ line fluxes\label{tab:properties}}
\tablewidth{0pt}
\tablehead{
\colhead{Properties} & \multicolumn{2}{c}{2MASS J16064794-1841437} & \multicolumn{2}{c}{2MASS J15582981-2310077} & \multicolumn{2}{c}{2MASS J16100541-1919362}
}
\startdata
Distance (pc, Gaia DR3) & \multicolumn{2}{c|}{156 $\pm$ 1} & \multicolumn{2}{c|}{141 $\pm$ 1} & \multicolumn{2}{c}{142 $\pm$ 1} \\
SpT & \multicolumn{2}{c|}{K9 \citep{Fang2023}} & \multicolumn{2}{c|}{M3 \citep{Manara2023}} & \multicolumn{2}{c}{M7 \citep{vanderplas2016}} \\
Stellar Mass (M$_{\odot}$) & \multicolumn{2}{c|}{0.56 \citep{Fang2023}} & \multicolumn{2}{c|}{0.144 \citep{Manara2023}} & \multicolumn{2}{c}{0.06 \citep{vanderplas2016}}\\
Log $\dot{M}_{acc}$ (M$_{\odot}$~yr$^{-1}$) & \multicolumn{2}{c|}{-9.42 \citep{Fang2023}} & \multicolumn{2}{c|}{-9.05 \citep{Manara2023}} & \multicolumn{2}{c}{\nodata} \\
Radial Velocity (km~s$^{-1}$) & \multicolumn{2}{c|}{-3.3 \citep{Fang2023}} & \multicolumn{2}{c|}{-6 \citep{Fang2023}} & \multicolumn{2}{c}{-8.01 \citep{ratzenbock2023b}} \\
Disk Inclination (deg) & \multicolumn{2}{c|}{55.5 $\pm$ 1 \citep{Ansdell2020}} & \multicolumn{2}{c|}{80.5 $\pm$ 32.7 \citep{Carpenter2025}} & \multicolumn{2}{c}{\nodata} \\
Position Angle (deg) & \multicolumn{2}{c|}{20 $\pm$ 1 \citep{Ansdell2020}} & \multicolumn{2}{c|}{53.2 $\pm$ 71 \citep{Carpenter2025}} & \multicolumn{2}{c}{\nodata} \\
Sco-Cen sub-group & \multicolumn{2}{c|}{$\beta$ Sco} & \multicolumn{2}{c|}{$\delta$ Sco} & \multicolumn{2}{c}{$\nu$ Sco} \\
Sub-group age (Baraffe, Myr) & \multicolumn{2}{c|}{4.6-7.6} & \multicolumn{2}{c|}{5.8-8.8} & \multicolumn{2}{c}{3.4-6.1} \\
Sub-group age (PARSEC, Myr) & \multicolumn{2}{c|}{6.9-8.4} & \multicolumn{2}{c|}{8.4-11} & \multicolumn{2}{c}{5.3-7.7} \\
FUV Flux (G$_0$) & \multicolumn{2}{c|}{13.5$^{+3.0}_{-3.5}$ \citep{Anania2025a}} & \multicolumn{2}{c|}{45$^{+55}_{-28}$ \citep{Anania2025a}} & \multicolumn{2}{c}{33$^{+41}_{-12}$ \citep{Anania2025a}} \\
Dominant FUV source & \multicolumn{2}{c|}{$\beta$~Sco} & \multicolumn{2}{c|}{$\delta$~Sco} & \multicolumn{2}{c}{$\nu$~Sco} \\
Apparent driving source & \multicolumn{2}{c|}{$\omega$~Sco (B1~V)} & \multicolumn{2}{c|}{$\delta$~Sco (B0.3~IV and B1-3~V binary)} & \multicolumn{2}{c}{$\omega$~Sco (B1~V)} \\
Projected 2D Distance (pc) & \multicolumn{2}{c|}{4.9} & \multicolumn{2}{c|}{1.7} & \multicolumn{2}{c}{3.86} \\
Estimated 3D Distance (pc) & \multicolumn{2}{c|}{6.2$^{+1.9}_{-0.7}$ \citep{Anania2025a}} & \multicolumn{2}{c|}{2.6$^{+2.5}_{-0.8}$ \citep{Anania2025a}} & \multicolumn{2}{c}{4.6$^{+2.7}_{-0.6}$ \citep{Anania2025a}} \\
Star--Bright front separation (au) & \multicolumn{2}{c|}{$\sim$470} & \multicolumn{2}{c|}{$\sim$220} & \multicolumn{2}{c}{$\sim$90} \\
H$_2$ mass-loss rate (M$_{\odot}$~yr$^{-1}$) & \multicolumn{2}{c|}{1.4 $\times$ 10$^{-9}$} & \multicolumn{2}{c|}{0.85 $\times$ 10$^{-9}$} & \multicolumn{2}{c}{0.4 $\times$ 10$^{-9}$} \\
\hline
\colhead{Line} & \colhead{PSF Flux (W~m$^{-2}$)} & \colhead{Extended Flux (W~m$^{-2}$)} & \colhead{PSF Flux (W~m$^{-2}$)} & \colhead{Extended Flux (W~m$^{-2}$)} & \colhead{PSF Flux (W~m$^{-2}$)$^a$} & \colhead{Extended Flux (W~m$^{-2}$)} \\
\hline
H$_2$ $\nu$0-0 S(1) 17.035~\micron{} & 4.24E-19 $\pm$ 2.36E-20 & 1.71E-18 $\pm$ 1.09E-19 & 4.41E-19 $\pm$ 8.96E-21 & 8.95E-19 $\pm$ 5.43E-20 & 3.38E-20 $\pm$ 3.39E-21 & 6.35E-19	$\pm$ 8.93E-20\\
H$_2$ S(2) 12.279~\micron{} & 1.94E-19 $\pm$ 5.10E-21 & 9.64E-19 $\pm$ 4.49E-20 & 2.88E-19 $\pm$ 1.48E-20 & 5.17E-19 $\pm$ 7.00E-20 & 4.41E-20 $\pm$ 4.80E-21 & 3.42E-19 $\pm$ 3.81E-20 \\
H$_2$ S(3) 9.665~\micron{} & 4.66E-19 $\pm$ 6.22E-21 & 2.34E-18 $\pm$ 2.59E-20 & 4.34E-19 $\pm$ 5.81E-21 & 2.07E-18 $\pm$ 6.65E-20 & 2.14E-20 $\pm$ 2.08E-21 & 3.66E-19 $\pm$ 2.61E-20 \\
H$_2$ S(4) 8.025~\micron{} & 2.60E-19 $\pm$ 1.80E-20 & 7.77E-19 $\pm$ 2.91E-20 & 2.26E-19 $\pm$ 1.26E-20 & 1.43E-18 $\pm$ 1.04E-19 & $<$ 1.40E-20 & $<$ 1.54E-19 \\
H$_2$ S(5) 6.909~\micron{} & 3.50E-19 $\pm$ 2.64E-20 & 1.43E-18 $\pm$ 5.86E-20 & 3.33E-19 $\pm$ 7.65E-21 & 1.64E-18 $\pm$ 1.44E-19 & $<$ 1.59E-20 &  $>$ 1.96E-19\\
H$_2$ S(6) 6.109~\micron{} & 2.24E-19 $\pm$ 3.15E-20 & 2.30E-19 $\pm$ 5.94E-20 & 2.29E-19 $\pm$ 3.54E-20 & 6.32E-19 $\pm$ 1.39E-19 & $<$ 3.25E-20 & $<$ 2.28E-19  \\
H$_2$ S(7) 5.511~\micron{} & $<$ 3E-19 & $>$ 8.5E-19 & 2.18E-19 $\pm$ 4.11E-21 & 5.49E-19 $\pm$ 1.05E-19 & $<$ 6.50E-20 & $<$ 4.45E-19 \\
\enddata
\tablecomments{For the star-forming region sub-groups, we list both the Baraffe \citep{Baraffe2015} and PARSEC isochronal ages \citep{Bressan2012,Chen2015,Marigo2017}; however, the latter were found to be the more robust among the two \citep{ratzenbock2023a,ratzenbock2023b}. Spectral types of the driving sources are from \cite{Negueruela2024} and \cite{Miroshnichenko2013}, respectively. 68\% confidence intervals are reported for the FUV fluxes and estimated 3D distances \citep{Anania2025a}. The 2D and 3D distances listed are to the apparent driving source. The star and bright-front separations (equivalently, dissociation-front radius) are reported as distances from the centroid to the bright-front outer edge towards the driving source. The separations are deconvolved from the PSF FWHM. Details on H$_2$ wind mass-loss rate calculations are in Section \ref{sec:h2_mass_loss_calculation}. The line fluxes are calculated within the PSF (diameter = 3$\times$FWHM, where FWHM varies with wavelength) and outside the PSF (extended emission = Full FOV flux - PSF flux). $^a$ For J1610-1919, the PSF flux is calculated within 2$\times$FWHM instead, accounting for the much closer dissociation front. The flux uncertainties include the 1$\sigma$ statistical uncertainty and 5\% calibration uncertainty. Uncertainties due to line blending from volatile emission are not considered.}
\end{deluxetable*}

Disks that are externally photoevaporating due to irradiation by nearby massive stars usually exhibit a distinctive ``cometary'' morphology in spatially resolved spectral lines tracing the unbound gas. Atomic and ionized lines trace the tail, while highly ionized lines (and radio free-free emission) resemble a cusp, oriented toward the irradiating massive star and tracing the ionization front \citep[e.g.,][]{Odell1993,Bally2000,Kim2016,Haworth2021,Aru2024}. The detection of such a morphology has thus far provided the most robust and direct evidence for externally photoevaporating disks (i.e., proplyds) in strong-FUV environments ($\gtrsim$10,000~G$_0$). In these high-FUV cases, the molecular emission typically traces the disk surface, if detected. In weak UV fields, instead, the molecular emission could be tracing a dissociation front offset from the disk, as was recently shown for two disks using MIRI H$_2$ line maps \citep{Bajaj2026}. Accordingly, in addition to the atomic and ionized lines, we also create H$_2$ line maps to search for this morphology. We construct spaxel-by-spaxel continuum-subtracted line intensity maps in \feii{} (bright lines at 5.340 and 17.936 \micron{}), \neii{}~12.813~\micron{}, \neiii{}~15.555~\micron{}, \arii{}~6.985~\micron{}, \hi{}~7-6, and H$_2$ S(1) to S(8).

We use the same method to create line maps as in \cite{Bajaj2025}. Briefly, for each spaxel, we consider approximately 15 wavelength points on either side of the peak line wavelength. We fit a straight line to the continuum and subtract it from the data; then fit a 1D Gaussian to the residual line profile to obtain the line intensity in that spaxel. We do not find any emission in the \feii{}, \arii{}, \hi{}, \neii{}, or \neiii{} lines that could be attributed to external PE based solely on their morphology \citep[see][for the \neii{} line maps of several Upper Sco sources]{Bajaj2026}. However, in H$_2$, we identify the cometary morphology toward three disks: 2MASS~J16064794-1841437, 2MASS~J15582981-2310077, and 2MASS~J16100541-1919362 (see Appendix \ref{sec:h2_line_maps} for H$_2$ line intensity maps of all sources). 

The H$_2$ line intensity maps from S(1) to S(5) for these three sources are shown in Figure \ref{fig:grid_maps}, along with \arii{}, \hi{}~7-6, \neii{}, \neiii{}, and \feii{}~17.936~\micron{} lines. Since the MIRI field of view (FOV) increases (and spatial resolution decreases) with increasing wavelength, we plot all maps on the same spatial scale, with each image centered on the centroid of the corresponding continuum at that wavelength. The clearest cometary morphology is seen towards J1558-2310, where all five H$_2$ lines trace the spatially extended cusp, with the para lines appearing weaker than the ortho lines (consistent with ortho-to-para ratio $>$ 1). Thus, the bright front is most clearly seen in the ortho lines, and we confirm that no extended emission is detected in the S(6) to S(8) lines. Similarly, the atomic and ionized lines do not exhibit any extended emission. Toward J1606-1841, we find the same overall results but with two key differences: (a) the bright front is located farther from the star/disk compared to J1558-2310, causing the emission to be truncated by the smaller FOV of the S(4) and S(5) lines, and (b) strong ionized emission centered on the continuum centroid is detected in nearly all the atomic and ionized lines shown. This emission is likely associated with photoionization driven by irradiation from the central star. Finally, toward J1610-1919, we find (a) the bright front to be closer to the source than both J1558-2310 and J1606-1841, (b) the bright front to be much flatter compared to the other sources, and (c) the emission to be weak enough to not be detected in S(4) and S(5) lines. The reason for its flat bright front is unclear; however, irradiation by multiple nearby stars would provide the simplest explanation.

Given that signs of external PE are detected toward only three sources and exclusively in the H$_2$ lines, we focus solely on these for the remainder of the paper. Their star/disk properties are provided in Table \ref{tab:properties} along with the H$_2$ line fluxes within the PSF and extended emission outside the PSF. In the next section, we assess whether the detected morphologies are indeed attributable to external PE.

\subsection{Confirmation of the `proplyd' nature} \label{sec:confirmation_proplyds}

\begin{figure*}
\centering
\includegraphics[width=\linewidth]{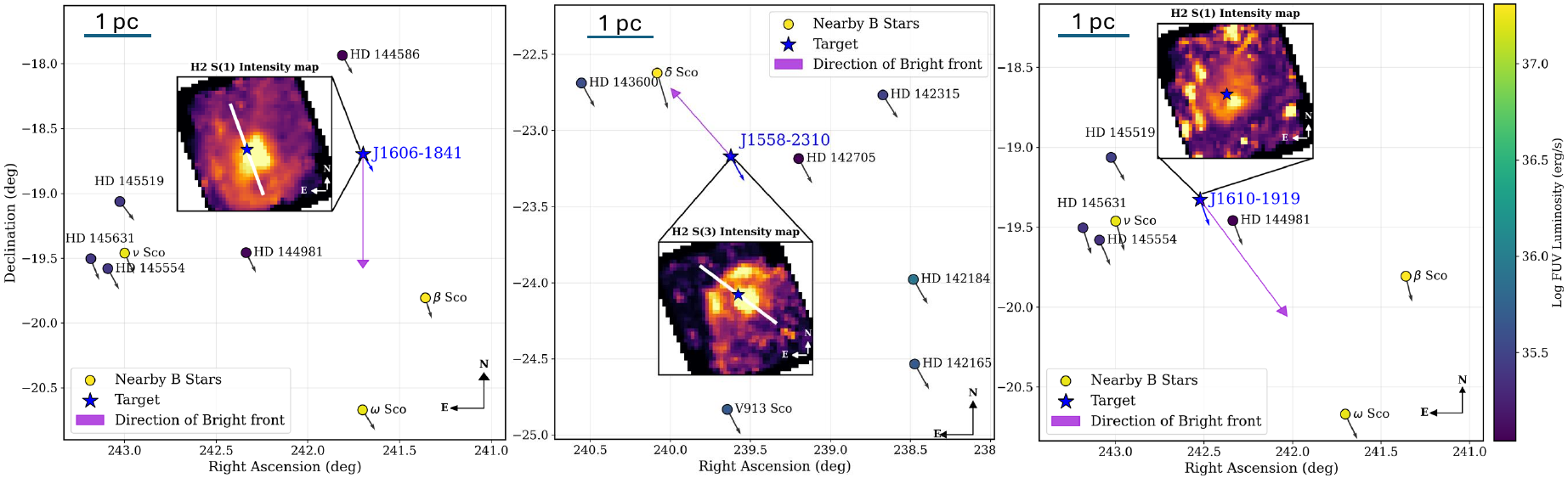}
\caption{The coordinates and FUV luminosities of the A/B stars near J1606-1841, J1558-2310, and J1610-1919 are taken from \cite{Anania2025a}. The positions of the blue stars in the inset images are based on the source coordinates and are therefore subject to the astrometric uncertainties of MIRI MRS ($\sim$0.3--0.5\arcsec). The white line passing through each source indicates the disk position angle (PA) as inferred from ALMA continuum observations; the disk PA for J1558-2310 is highly uncertain. Arrow lengths represent proper motions in units of mas~yr$^{-1}$ and are scaled up by a factor of 150. The purple arrows are made by connecting the stellar position (blue star) to the bright front in the insets.}
\label{fig:Direction_confirmation}
\end{figure*}

Since the observed H$_2$ morphologies could have alternative origins, such as bow shocks resulting from differential motion between the star-disk system and the surrounding gas \citep[e.g.,][]{Peri2012}, we plot the positions and proper motions of the star-disk systems along with nearby B-type stars in Figure \ref{fig:Direction_confirmation}. The data on the positions, proper motions, and FUV luminosities of the nearby B-type stars are taken from \cite{Anania2025a}. A key fundamental property of proplyds is that their ionization front points directly toward the irradiating source responsible for the dominant FUV photons. From Figure \ref{fig:Direction_confirmation}, in all three cases, the molecular dissociation fronts point toward one of the brightest nearby FUV sources: for J1606-1841 and J1610-1919 this is $\omega$~Sco, while for J1558-2310, this is $\delta$~Sco.

For J1606-1841 and J1558-2310, the molecular dissociation fronts point in directions distinct from the proper motion vectors of the star-disk systems. Moreover, J1558-2310 is moving slightly faster relative to the average $\delta$~Sco sub-cluster motion \citep[by $\sim$0.4~mas~yr$^{-1}$ in both RA and DEC,][]{ratzenbock2023a}, and J1606-1841 is moving slower relative to the $\beta$~Sco sub-cluster \citep[by $\sim$0.8 and 0.7~mas~yr$^{-1}$ in RA and DEC, respectively,][]{ratzenbock2023a}. For J1558-2310 and J1606-1841, this means their H$_2$ bow shocks would be pointing towards and against the direction of their proper motion, respectively, which is inconsistent with the direction of the observed H$_2$ dissociation fronts, indicating that their H$_2$ morphologies are inconsistent bow shocks. For J1610-1919, while the direction of its motion and dissociation front is similar, it is moving slower than $\nu$~Sco, the subgroup it belongs to \citep[by $\sim$0.5~mas~yr$^{-1}$ in RA,][]{ratzenbock2023a}, suggesting the origin of its extended H$_2$ emission against a bow shock as well.

Another possibility could be that of H$_2$ tracing an internally driven wind excited by external radiation \citep[e.g.,][]{Clarke2015}. Recently, it was found that the detection fraction of extended H$_2$ winds drops with decreasing mass accretion rate, with detection rates below 50\% for $\dot{M}_{\rm acc}\lesssim10^{-9}M_{\odot}\mathrm{yr^{-1}}$ \citep{Bajaj2026}. Moreover, the detection fraction drops below 30\% for disks in the Upper Sco region. Given that none of the three disks are particularly high accretors (Table \ref{tab:properties}), the likelihood of the observed H$_2$ emission originating in internal winds is low. Furthermore, the four Upper Sco disks with H$_2$ emission attributed to internal winds by \citet{Bajaj2026} experience a higher mean FUV field ($\sim43~G_0$) than the three disks studied here ($\sim31~G_0$). If the H$_2$ morphology observed in our sources resulted from external illumination of internally driven winds, similar morphologies would be expected in these more strongly irradiated systems. Their absence disfavors this interpretation and instead supports an externally driven photoevaporative origin for the winds studied here. Following this confirmation, we also include the spectral types and distances of the driving sources in Table \ref{tab:properties} along with the estimated separation between the stellar centroid and the molecular dissociation front in the direction of the driving source.

\begin{figure*}
\centering
\gridline{
    \raisebox{-\height}{\fig{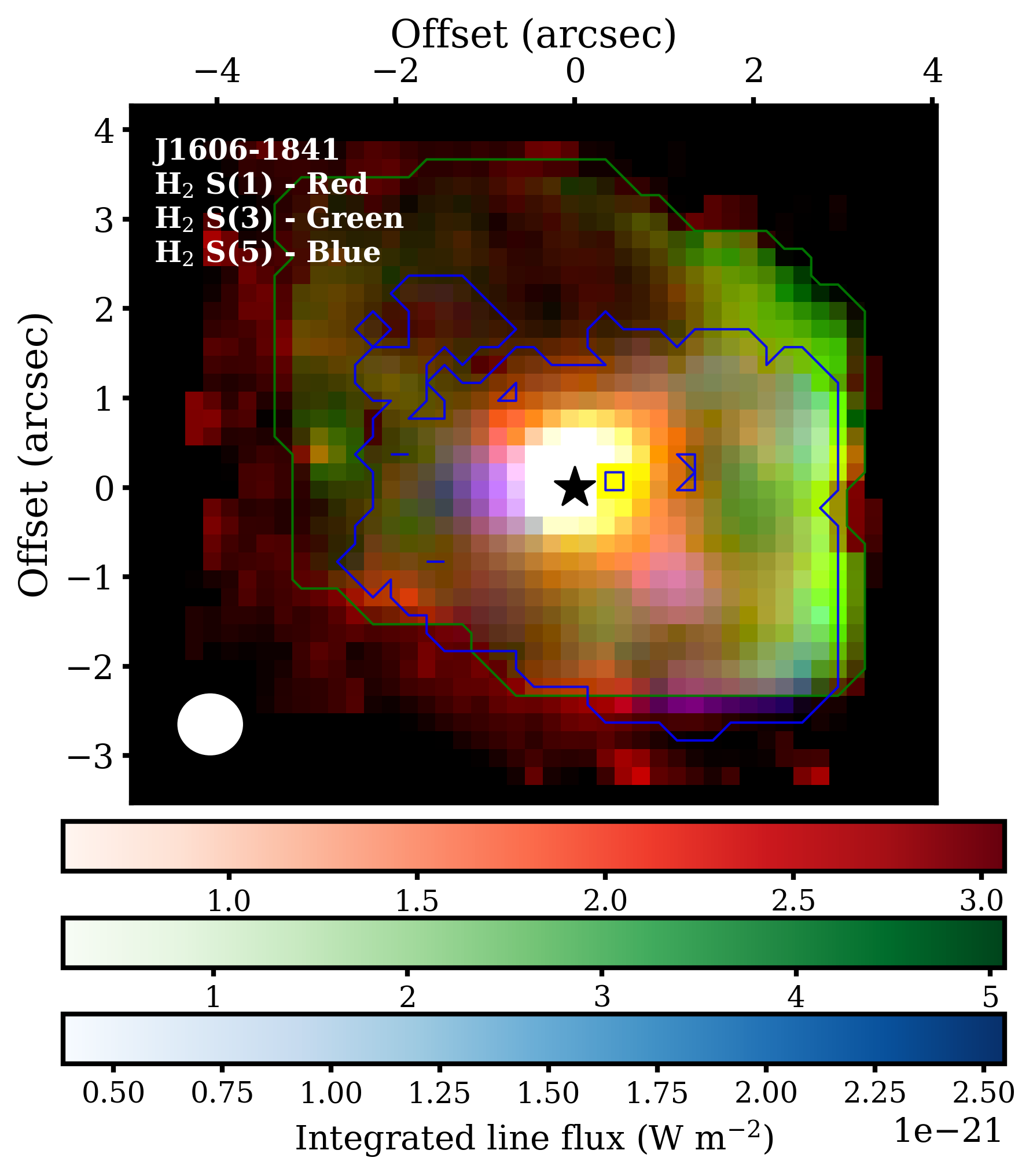}{0.32\textwidth}{(a) J1606-1841}}
    \raisebox{-\height}{\fig{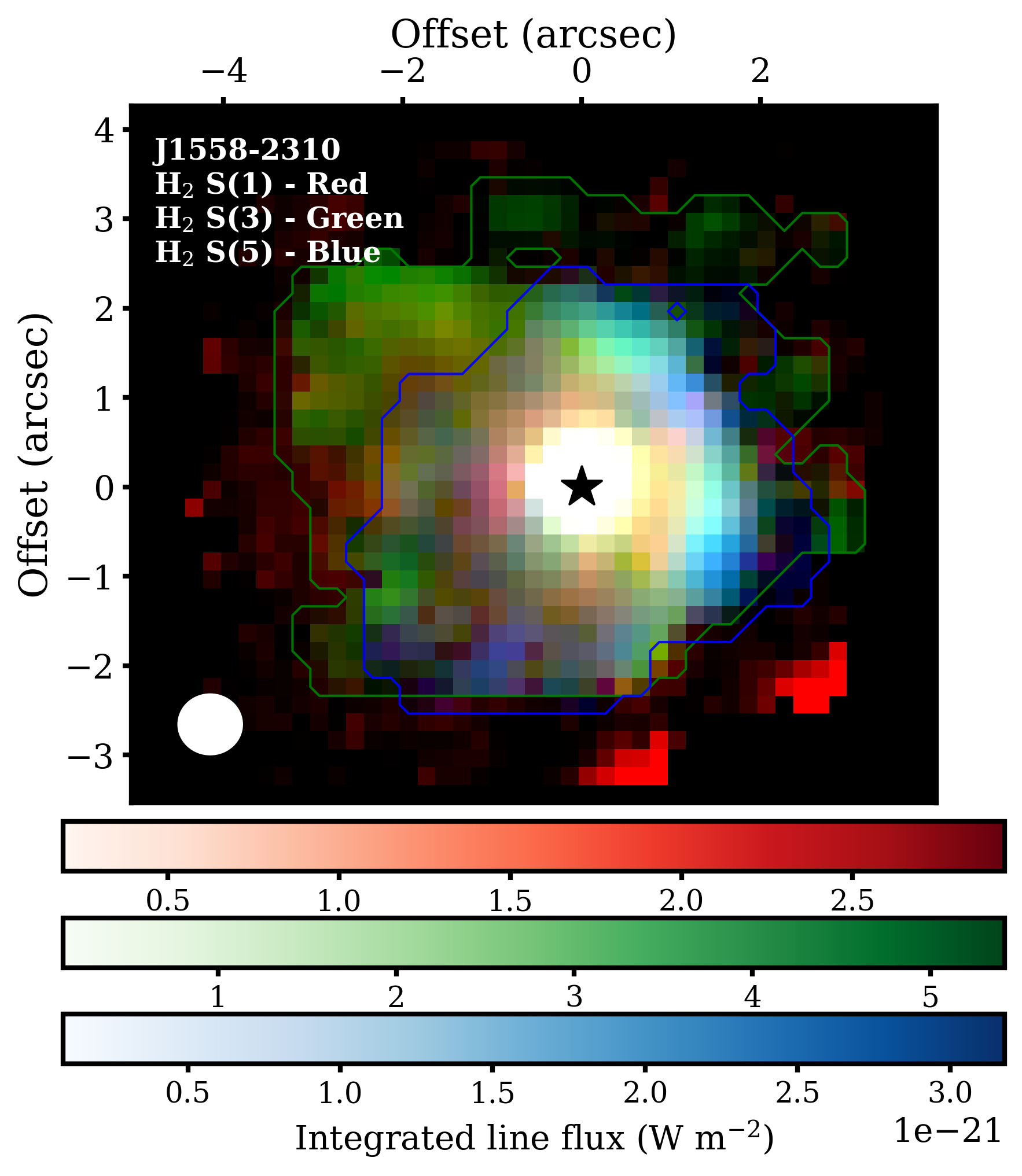}{0.32\textwidth}{(b) J1558-2310}}
    \raisebox{-\height}{\fig{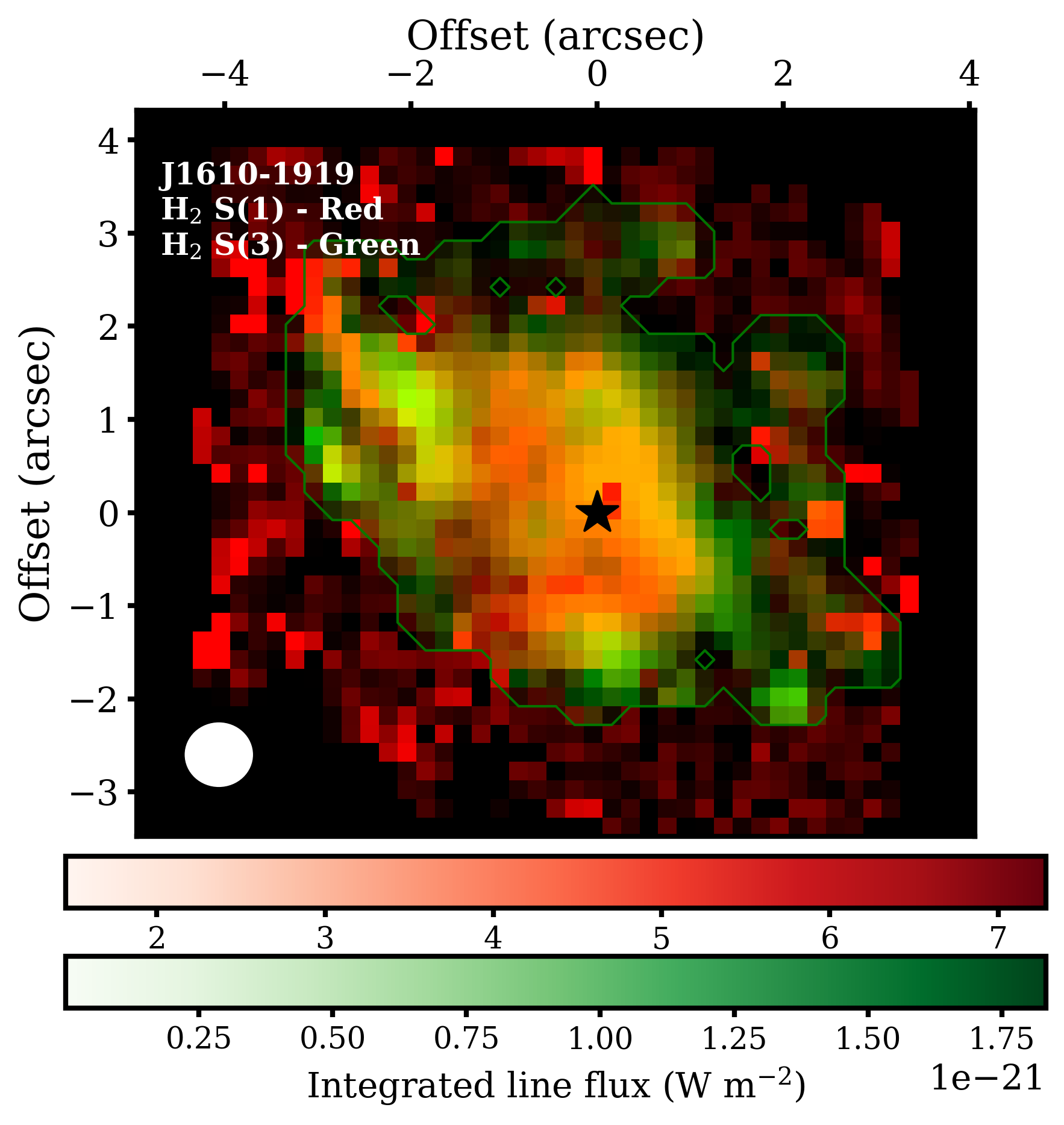}{0.316\textwidth}{\vspace{0.35cm}(c) J1610-1919}}
}
\caption{RGB composite images constructed from the H$_2$ line intensity maps, with S(1) at 17.035~\micron{} shown in red, S(3) at 9.665~\micron{} in green, and S(5) at 6.909~\micron{} in blue. For J1610-1919, only the red and green channels are plotted, as the blue channel (S(5)) only adds noise to the image. Before generating the RGB images, the higher-resolution S(3) and S(5) maps were convolved to the lower spatial resolution of the S(1) map. Each channel is displayed using a linear intensity scale with lower and upper cuts set to the 5th and 95th percentiles of the corresponding line intensity distribution, respectively. The black star is plotted at the centroid of the continuum near the H$_2$~S(1) line. The blue and green contours mark the field of view containing valid intensity values in the S(5) and S(3) line, respectively.}
\label{fig:H2_overplot}
\end{figure*}

The term proplyd is defined as ``a circumstellar disk with an externally driven photoevaporative wind composed of a photodissociation region and an exterior ionization front with a cometary morphology'' \citep{Odell1993,Winter2022}. Since we do not detect an ionization front in the data (see atomic and ionized line maps in Figure \ref{fig:grid_maps}) and instead find an exterior H$_2$ photodissociation front, we choose to refer to these objects as ``molecular proplyds'', clearly highlighting this key distinction. However, given that these objects reside in very low FUV environments ($<100~G_0$), the ionization fronts likely lie beyond the FOV of current MIRI observations. As such, the use of the term ``molecular proplyds" may only be temporary until wider FOV observations are carried out to verify the existence of detectable ionization fronts.

Finally, we show WISE three-color images of $\delta$~Sco, $\omega$~Sco, $\beta$~Sco, and $\nu$~Sco, the dominant FUV sources near the three molecular proplyds, in Appendix \ref{sec:bow_shock_WISE}. In these images, $\delta$~Sco clearly exhibits a bow shock consistent with the direction of its motion. \cite{Peri2012} first reported this and suggested that $\delta$~Sco is passing through the region, consistent with its larger and slightly offset proper motion compared to nearby stars, which, however, also carries larger uncertainties. The WISE images additionally reveal a potential bow shock associated with $\omega$~Sco. While it is not as pronounced as that of $\delta$~Sco, it appears similar to other bow shocks, e.g., HIP~77391, identified in the survey of \cite{Peri2012}. This suggests that $\delta$~Sco (irradiation source for J1558-2310) and potentially $\omega$~Sco (one of the irradiation sources for J1606-1841 and J1610-1919) are runaway stars, confirming that external PE in this region is time-variable.

\subsection{H$_2$ Rotational diagram and mass loss estimation} \label{sec:h2_mass_loss_calculation}

\begin{figure*}
\centering
\includegraphics[width=\linewidth]{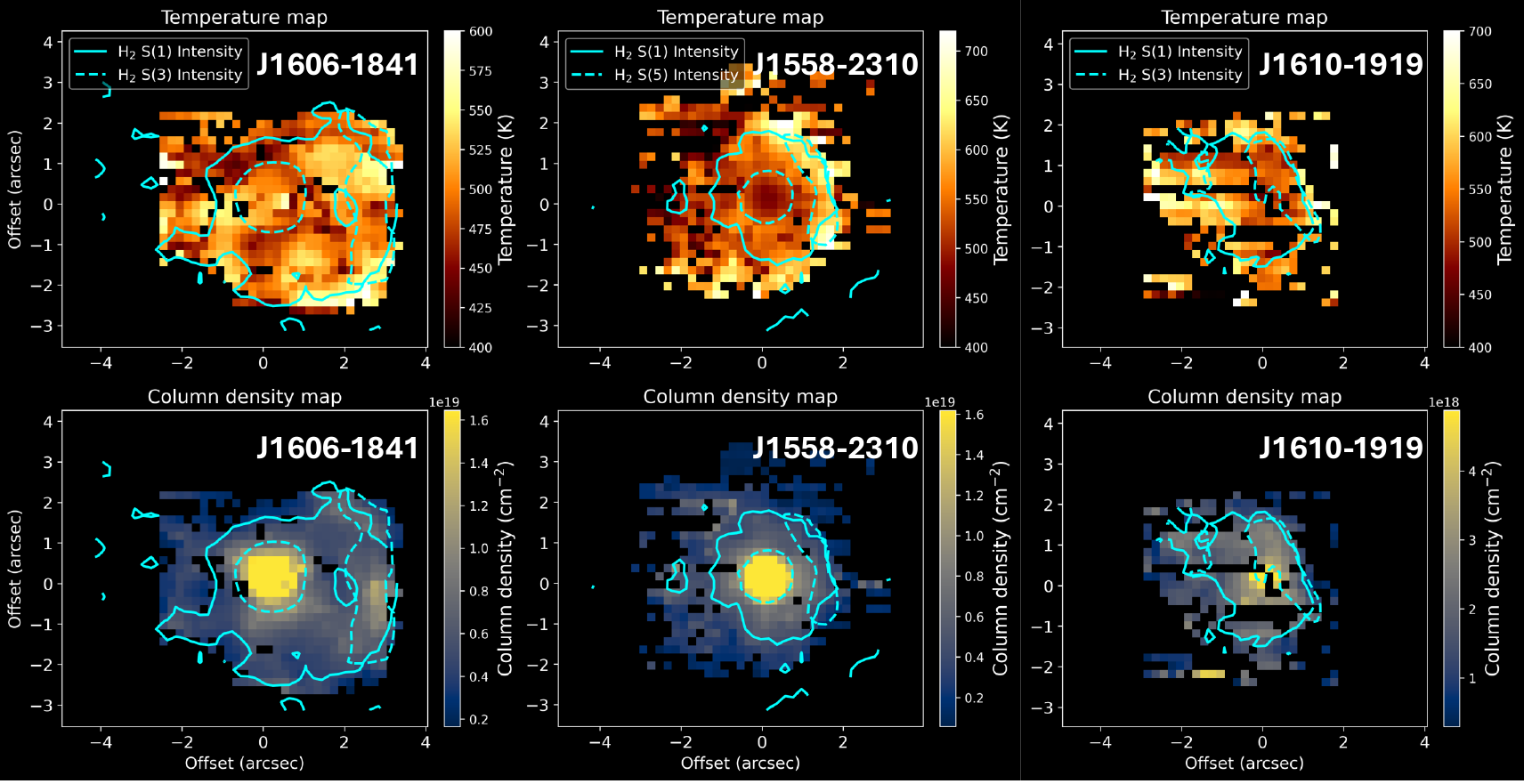}
\caption{Pixel-by-pixel best-fit temperature (top) and column density (bottom) maps derived from rotational diagram fits to the H$_2$~$\nu$=0-0 S(1)-S(5) lines toward J1606-1841, J1558-2310, and J1610-1919. Cyan contours denote 3$\sigma$ levels of the H$_2$ S(1) (solid) and S(5) or S(3) (dashed) line intensities, where $\sigma$ is calculated iteratively using \texttt{astropy sigma\_clip} on the maps shown in Figure~\ref{fig:grid_maps}. All line maps were convolved to the resolution of H$_2$~S(1) and registered to the nearby continuum centroid before constructing the rotational diagrams.}
\label{fig:temp_dens_maps}
\end{figure*}

To assess the impact of external PE on the masses and lifetimes of these disks, an estimation of the wind mass-loss rate is necessary. This can be obtained through H$_2$ rotational diagrams, which provide the temperature and column density of the emitting gas under the assumption of local thermodynamic equilibrium (LTE) conditions. The non-detection of the $\nu$1-1 H$_2$ transitions in the MIRI spectra, along with the non-detection of $\nu$0-0 S(6)-S(8) at the dissociation front (towards J1558-2310 and J1610-1919 where the FOV allows their detection) suggests that the observed H$_2$ emission is mostly thermally excited. Rotational diagrams created using the total line fluxes summed over the entire FOV (those reported in Table \ref{tab:properties}) would also include any emission from the disk that is not necessarily associated with the externally driven photoevaporative wind. Therefore, we instead perform a spaxel-by-spaxel rotational diagram fit and create temperature and column density maps to isolate the external photoevaporative wind emission.

Since the S(1) to S(5) lines fall in different MIRI channels and subbands, they have different FOVs and resolution. Accordingly, we first convolve all line maps to the resolution of the S(1) line (longest wavelength, poorer resolution) and regrid the emission to the same pixel size and FOV as S(1). Any pixels lying outside the FOV of the original unconvolved line maps are masked, and all maps are registered such that the position of their continuum centroids matches that near the S(1) line. This assumes that the continuum centroid at each wavelength traces the same physical object (the star/disk) and introduces significantly lower uncertainties than the astrometric calibration of MIRI, which can produce offsets as large as 0.5 arcsec ($\sim$two pixel widths at S(1) resolution) between two different lines. RGB composite images created using these convolved H$_2$ line maps are shown in Figure \ref{fig:H2_overplot}.

We then consider the S(1) to S(5) line fluxes, including the para lines, in each pixel and fit a single temperature component assuming an ortho-to-para ratio of 3. An example fit at the dissociation fronts can be found in Appendix \ref{sec:example_rotational_diag}. In every pixel, a fit is only performed when three or more lines are detected, as there are two free parameters. We do not correct the line fluxes for extinction, as the observed extinction toward these sources results in flux variations significantly smaller than the calibration uncertainties \citep[A$_V$ $\sim$ 0.7~mag;][]{Rilinger2021,Fang2023}. The resulting temperature and column density maps are shown in Figure \ref{fig:temp_dens_maps} for all three sources, with H$_2$ intensity contours overplotted. Towards J1558-2310 and J1606-1841, we find significantly elevated temperatures near the photodissociation front compared to the star+disk position and, inversely, decreased column densities. At least for J1558-2310, this trend is already apparent in the RGB composite image (Figure \ref{fig:H2_overplot}), where the S(5) line (in blue) is the brightest at the photodissociation front, suggesting higher temperatures there, while for J1606-1841, the S(5) line emission is partially truncated by the FOV.

Finally, to estimate the wind mass loss rates, we convert the column densities to H$_2$ masses in every pixel and sum the masses in the bright rim near the photodissociation front, e.g., the rim seen in H$_2$~S(5) towards J1558-2310 in Figure \ref{fig:grid_maps} (see Section \ref{sec:example_rotational_diag} and Figure \ref{fig:mass_region_highlighted} for the exact regions towards all three sources). The selected rim sizes provide a lower limit on the total wind mass, assuming that some mass may be invisible under the observed conditions. In case of a spherically symmetric wind, a correction factor of 2$\pi$/$\phi$ would need to be applied to the total mass, where $\phi$ is the angle projected by the rim to the disk center. For our selections (Figure \ref{fig:mass_region_highlighted}), this factor would range from $\sim$2 to $\sim$4. However, given the importance of the 3D disk-B-star geometry in weak fields (discussed later in Section \ref{sec:discuss_fried_comparison}), the wind is highly unlikely to be spherically symmetric, and the correction factor would be at most $\pi$/$\phi$, i.e., $\lesssim$2.

We estimate H$_2$ masses of 4.3~$\times$~10$^{-7}$~M$_{\odot}$, 1.1~$\times$~10$^{-7}$~M$_{\odot}$, and 3.1~$\times$~10$^{-8}$~M$_{\odot}$ toward J1606-1841, J1558-2310, and J1610-1919, respectively, and crossing timescales of 308~yr, 126~yr, and 71~yr. The latter is estimated by dividing the PSF-deconvolved width of the rim by a wind velocity of 2.5~km~s$^{-1}$ (see Equation \ref{eq:1}), i.e., the sound speed for H$_2$ gas at the observed temperatures of 600-700~K, since we expect the gas to be supersonic by the H-H$_2$ transition \citep{Haworth2018}. 

\begin{equation} \label{eq:1}
    \dot{M}_{wind} = M_{H_2-rim} \div (width_{rim}/c_s)
\end{equation}

Finally, dividing the rim mass by the crossing timescale provides wind mass loss rates of 1.4~$\times$~10$^{-9}$~M$_{\odot}$~yr$^{-1}$, 0.85~$\times$~10$^{-9}$~M$_{\odot}$~yr$^{-1}$, and 0.4~$\times$~10$^{-9}$~M$_{\odot}$~yr$^{-1}$ for J1606-1841, J1558-2310, and J1610-1919, respectively. Interestingly, the mass-loss rates directly correlate with the dissociation front radii (listed in Table \ref{tab:properties}), as expected for similar FUV fluxes (T. Haworth, priv. comm.).

\subsection{Theoretical mass loss rate predictions} \label{sec:comparison_theoretical_pred_allsources}

\begin{figure*}
    \centering
    \includegraphics[width=\linewidth]{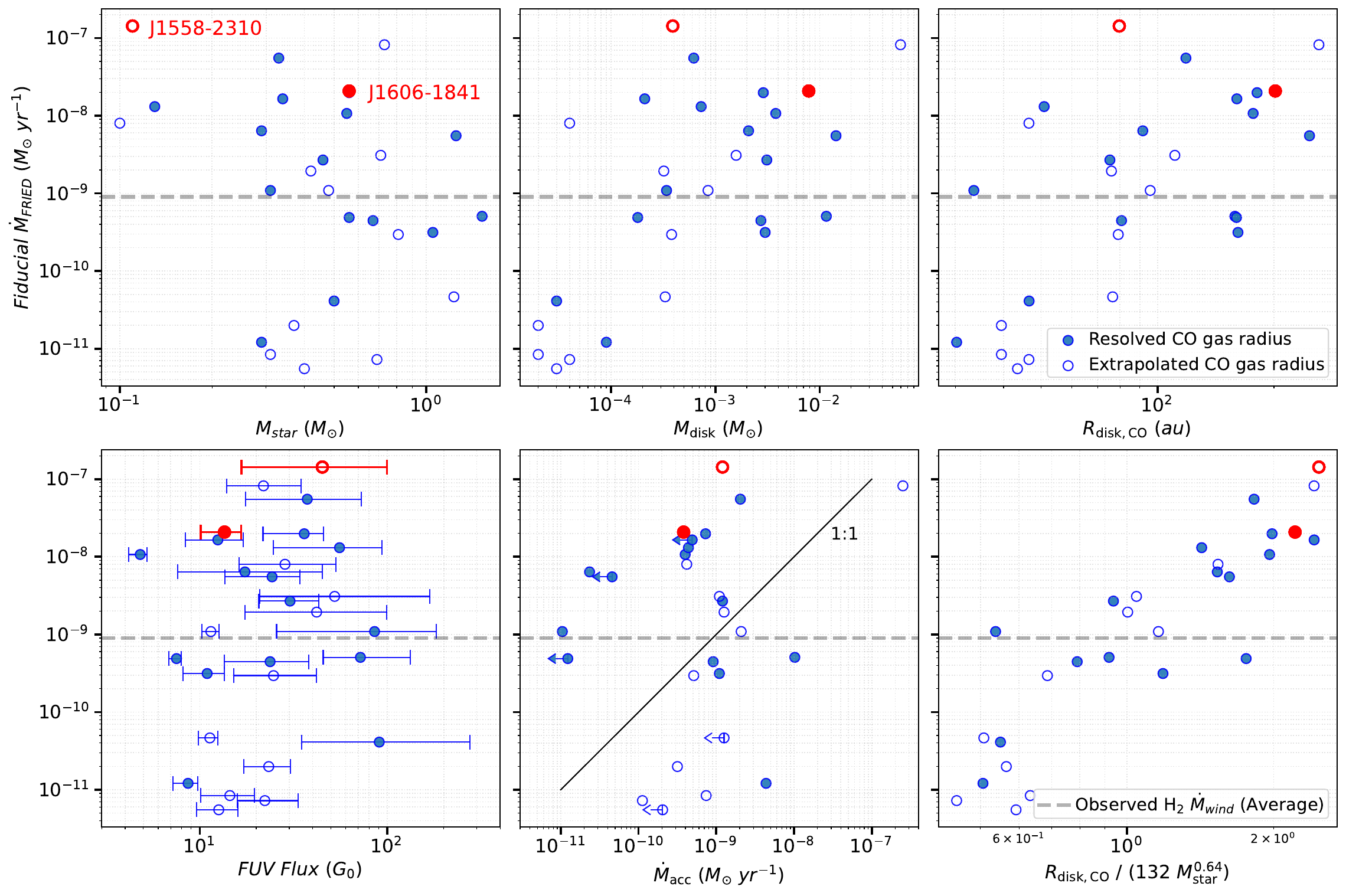}
    \caption{Fiducial ($f_{\rm PAH,d}=1$) \textsc{FRIEDv2} mass loss rate estimates for the 28 Upper Sco disks (primary sample) observed with MIRI, plotted against the input stellar and disk parameters, the mass accretion rate, and the unitless quantity $R_{\rm disk,CO}/(132 \times M_{\rm star}^{0.64})$. Details of the mass-loss rate calculations are provided in Section~\ref{sec:comparison_theoretical_pred_allsources}. The two molecular proplyds are highlighted in red. J1610-1919 is not included (along with other very-low-mass disks) since \textsc{FRIEDv2} grids only cover stellar masses $>$0.1~M$_{\odot}$. Solid circles denote sources with directly measured gas disk radii \citep{Trapman2025,Zallio2026a}, while hollow circles indicate sources with indirectly estimated (and hence more uncertain) gas disk radii (see Section \ref{sec:comparison_theoretical_pred_allsources} for details).}
    \label{fig:FRIED_4d_comparison}
\end{figure*}

To derive predicted mass-loss rates for the molecular proplyds, as well as the other Upper Sco objects with MIRI/MRS data, we use the \textsc{FRIEDv2} grid of mass-loss rates \citep{Haworth2023}. The grid provides mass-loss rates tabulated at various values of the local FUV field (measured in units of $G_0$), stellar mass $M_*$, disk outer radius $R_{\rm disk}$, and gas mass/surface density. Since the \textsc{FRIEDv2} grid covers only stellar masses $\geq$ 0.1~M$_{\odot}$, we exclude the five very-low-mass/brown-dwarf disks from this analysis and focus on the primary sample of 28 disks.
To enable smoother interpolation, we first convert the disk mass to surface density at 1~au, since the grid of models was constructed by varying this quantity \citep[c.f.][]{Sellek2020}. The interpolation is performed using the \texttt{scipy} LinearNDInterpolator in \textbf{4D}, taking the logarithm of each quantity. For $\Sigma_1$ values below the range covered by the grid, rates are extrapolated proportionally to $\Sigma_1$ following the approach of \citet{Sellek2020}. As our default, we adopt the fiducial ``grain growth'' $f_{\rm PAH,d}=1$ grid, which typically yields the highest predicted mass loss rates among the six combinations of microphysical parameters included in \textsc{FRIEDv2}.

For each source, the stellar mass was taken from \citet{Manara2023} or \citet{Fang2023} and the FUV field was calculated using the method of \citet{Anania2025a}. The disk mass $M_{\rm disk}$ was estimated as $100\times$ the dust mass inferred from the ALMA band 7 continuum flux \citep{Carpenter2025} using Equation (1) from \citet{Ansdell2016} \citep[see also][]{Hildebrand1983} at a mean distance of 145~pc. This assumes optically thin blackbody emission at 20\,K with an opacity of $3.4\,\mathrm{cm^2\,g^{-1}}$ \citep{Beckwith1990}. Where available, the disk radius was taken as the 90\% \textsuperscript{12}CO flux radius from either \citet{Trapman2025} or \citet{Zallio2026a}; in case two values were available, we preferred whichever was smaller for a conservative estimate. In cases where no gas radius was available, we estimated the radius using $R_{\rm disk}=509\,\mathrm{au} \left(M_{\rm disk}/M_\odot\right)^{0.236}$, derived from a fit to those sources for which both the mm flux and CO radius were available. 

In Figure~\ref{fig:FRIED_4d_comparison}, we plot the predicted \textsc{FRIEDv2} wind mass loss rates against the four input star/disk parameters ($M_{\rm star}$, $M_{\rm disk}$, $R_{\rm disk}$, and FUV flux), against the mass accretion rates from \citet{Manara2023}, \citet{Fang2023} or \citet{Empey2026}, and against $R_{\rm disk,CO}/(132 \times M_{\rm star}^{0.64})$, where the relation $R_{\rm disk,CO} = 132 \times M_{\rm star}^{0.64}$ was obtained by fitting a straight line between $\log_{10}M_{\rm star}$ and $\log_{10}R_{\rm CO/90}$, homogeneously calculated for 33 Upper Sco disks \citep{Zallio2026a,Zallio2026b}. We recover the expected trends: the wind mass loss rate increases with decreasing stellar mass, increasing disk mass, and increasing disk radius, though with considerable scatter. No obvious correlation is found with FUV flux, likely indicating that between 10 and 100~$G_0$, the star and disk parameters determine how strongly a disk is affected by external radiation. Similarly, no correlation is seen with mass accretion rate, though more disks have predicted wind mass-loss rates exceeding their accretion rates than falling below. 

The strongest correlation is seen with the unitless quantity R$_{disk,CO}$/(132~M$_{star}^{0.64}$), clearly showing that for a given stellar mass and for the explored FUV fluxes, the gas disk size (equivalently, the gravitational binding of the outer disk gas) determines whether significant external PE occurs: disks with R$_{disk,CO}$ $>$ 132~$\times$~M$_{star}^{0.64}$ have predicted external PE rates $\gtrsim$10$^{-9}$~M$_{\odot}$~yr$^{-1}$ (the average observed H$_2$ wind mass-loss rate). Finally, the highest predicted mass-loss rate is for J1558-2310, consistent with the clear proplyd morphology detected toward this source. J1606-1841 is also among the three most massive winds when considering sources with properly estimated disk radii. This, along with the fact that neither disk is a particularly high accretor \citep[thus they are less likely to have strong internal winds;][]{Bajaj2026}, demonstrates that it is physically entirely reasonable that these disks have extended H$_2$ emission dominated by external PE. Possible reasons for why other sources with high predicted mass-loss rates do not exhibit proplyd emission morphology in H$_2$ are discussed in Section~\ref{sec:discuss_fried_comparison}. 

Figure \ref{fig:FRIED_4d_comparison} shows that the FRIED fiducial mass-loss rates can be up to two orders of magnitude larger than the observed H$_2$ rates. This is likely due to a combination of the following effects: (a) the FRIED grid assumes isotropic irradiation on the disk, equally affecting the disk surface and disk edge. However, in such low FUV fields, the 3D geometry between the disk and the nearby massive stars becomes important and can reduce the theoretical estimates by 5-10 times \citep[see Section \ref{sec:discuss_fried_comparison}][]{Adams2004,Winter2019}. (b) Similarly, the model assumes an axisymmetric wind, whereas the observed estimates are derived from a region subtending an angle between $\pi$/2 and $\pi$ at the center (see Section \ref{sec:h2_mass_loss_calculation} and Figure \ref{fig:mass_region_highlighted}). This makes the model estimates systematically 2-4 times larger than those observed. (c) The mass loss rates plotted in Figure \ref{fig:FRIED_4d_comparison} are for fiducial values of PAH abundance and dust grain size and abundance; changing these parameters can reduce the predicted mass loss rates up to 20 times (see Figure \ref{fig:PAH_abundance_massloss} for an example). (d) The disk mass/radius is not always well-known (see hollow circles in Figure \ref{fig:FRIED_4d_comparison}). (e) Lastly, it is also possible that only a fraction of the outflowing gas is detectable in H$_2$ emission.

\section{Discussion} \label{sec:discussion}

In this section, we discuss the potential of spatially resolved mid-IR H$_2$ emission as a tracer of external PE in low-UV environments (Section \ref{sec:h2_tracer_proplyds}), the impact of external PE on the dispersal of these low-UV disks in comparison with internal dispersal processes (Section \ref{sec:discussion_low_mass_stars_impact_dispersal}), and finally why only three molecular proplyds are found and what this implies for external PE in such a region (Section \ref{sec:discuss_fried_comparison}).

\subsection{Comparison with archetypical proplyds: H$_2$ as the key tracer of low-FUV proplyds} \label{sec:h2_tracer_proplyds}

To date, most studies of spatially resolved externally photoevaporating disks have focused on the Orion Nebula Cluster (ONC), where proplyds were first unambiguously identified using the \texttt{Hubble Space Telescope} \citep{Odell1993}. Proplyds in this region experience FUV fluxes ranging from $<1000~G_0$ in the outskirts \citep[e.g.,][]{Pabst2021,Diaz-Berrios2024} to $>10^6$~G$_0$ in the innermost regions \citep[e.g.,][]{Aru2024}, with the classic proplyds near the Trapezium facing 10$^4$-10$^6$~G$_0$. Morphologically, these proplyds exhibit the cometary structure described earlier. Recently, \cite{Amiot2026} spatially resolved many ONC proplyds using NIRCam and studied the gas morphologies in \feii{}, Pa~$\alpha$, and H$_2$~1-0~S(1), classifying them into three types. In all three types, the H$_2$ emission traces the dissociation front at or near the disk surface, while the ionization front (Pa~$\alpha$) is found near the disk surface (Type~I), farther from the disk surface (Type~II), or absent (Type~III), depending on whether the ionizing Extreme Ultra-Violet (EUV) reaches the disk and the role it plays relative to the FUV.

J1606-1841, J1558-2310, and J1610-1919 experience FUV fluxes $<100~G_0$, making them the lowest G$_0$ targets with unambiguous direct evidence of externally driven photoevaporative winds. Their gas morphologies are in stark contrast to those seen in the ONC. Rather than a dissociation front near the disk surface, the dissociation fronts lie at hundreds of au from the disk (see Table~\ref{tab:properties}), placing them outside the three types characterized by \cite{Amiot2026}. This positions spatially resolved H$_2$ emission as a unique tracer of external PE in low $G_0$ environments \citep[$<100~G_0$ and likely up to 1000~G$_0$;][]{Booth2026}, opening a window to study its effects in nearby low-mass star-forming regions where much of current planet formation research has been concentrated. 

We do not detect extended I-fronts in any atomic or ionized lines, similar to Type~III ONC proplyds where EUV does not reach the star-disk system. However, at least in the case of $\delta$~Sco, the Str\"{o}mgren radius estimated by \citet{MiretRoig_2022} is 5.5-7.9\,pc, while the estimated separation of J1558-2310 is $2.6_{-0.8}^{+2.5}$\,pc, which would place it inside the Str\"{o}mgren sphere, meaning that EUV would impinge upon the flow.
Therefore, alternatively, these proplyds could be similar to the FUV-driven Type~II proplyds, but where the relatively low ionizing flux of a B star leads to a very distant I-front \citep{Johnstone1998}, possibly beyond the MIRI FOV, naturally accounting for the non-detection. Wider-field observations targeting ionized lines would be required to determine if there is a detectable I-front in the flow.

Recently, \cite{Rogers2025} conducted 1-5~\micron{} spectroscopy of 42 accreting pre-main-sequence stars in NGC~3603, where the disks face FUV fluxes $\gtrsim10^{5}~G_0$, and reported non-detections of rotational and rovibrational H$_2$ lines in that wavelength range. They argued that the H$_2$ emission, whether from internal winds or external PE, is likely ionized or entirely photodissociated by the intense external radiation in such a high UV environment. At lower G$_0$, \cite{Booth2026} analyzed MIRI observations of seven proplyds in NGC~1977 experiencing 10$^3$-10$^5$~G$_0$. They detected rotational H$_2$ emission lines towards all but the most highly irradiated disk ($\sim$10$^{4.8}$~G$_0$). Interestingly, they detected extended H$_2$ emission tracing an externally driven PE wind towards the disk facing the lowest FUV flux ($\sim$10$^{3.1}$~G$_0$). Combined with the three molecular proplyds at $<$100 G$_0$ reported in this study, this suggests a transition in the wind conditions/chemistry at $\sim$1000~G$_0$, such that above this threshold, H$_2$, when detected, traces the dissociation front at/near the disk. However, below this threshold, the dissociation front lies tens to hundreds of au away from the disk, leading to extended H$_2$ emission. Importantly, this transition makes H$_2$ the best-known tracer of external PE in weak ($\lesssim$~1000~G$_0$) UV environments. However, since this inference is based on only four proplyds, a larger MIRI sample analysis is required to further test this trend, ideally in the 100-1000~G$_0$ range.

\subsection{The impact of external PE on disk dispersal: low-mass stars/disks are significantly more impacted} \label{sec:discussion_low_mass_stars_impact_dispersal}

The proplyds detected and characterized to date belong to star-forming regions younger than 5~Myr \citep[see Table~1 of][for a compilation]{Winter2022}. J1606-1841, J1558-2310, and J1610-1919, by contrast, belong to the older Sco-Cen association, in sub-regions $\delta$~Sco, $\beta$~Sco, and $\nu$~Sco, respectively, with median age $>$5~Myr (Table~\ref{tab:properties}), making them the oldest externally photoevaporating disks discovered thus far. Their advanced ages, combined with the low FUV fluxes they experience, present an interesting case for understanding the impact of external PE on their disk lifetimes and how it compares with the internal processes.

If we assume that all three disks formed with masses at the gravitationally unstable limit (0.1~$M_{\rm star}$) and experienced a constant external photoevaporative mass loss rate equal to those calculated here, J1606-1841, J1558-2310 and J1610-1919 would require $\sim$34, $\sim$12.5~Myr, and $\sim$15~Myr, respectively, to reach their current disk masses. Under these assumptions, all three disks could have sustained the current UV levels throughout their evolution without running into the ``proplyd lifetime problem" reported in the ONC \citep[where the observed mass loss rates could disperse the entire disk at timescales much shorter than the disk age, e.g.,][]{Bally1998b,Henney1999}. For J1606-1841, if its initial disk mass was indeed close to the gravitationally unstable limit, then its disk dispersal thus far was likely dominated by internal winds and accretion rather than external PE. For J1558-2310 and J1610-1919, however, external PE could have been the dominant dispersal mechanism throughout their lifetimes. We also note that \cite{Anania2025b} found it easier to reproduce the current gas disk sizes of 7 out of 10 Upper Sco systems with an initial disk mass closer to 0.01~$M_{\rm star}$ using their external PE + viscous disk evolution models. For J1606-1841, such a low initial disk mass is not possible since its current disk mass is estimated at 0.014~$M_{\rm star}$. For J1558-2310 and J1610-1919, assuming an initial disk mass of 0.01~$M_{\rm star}$  with a constant external PE mass loss rate of 0.85 and 0.4~$\times$~10$^{-9}$~M$_{\odot}$~yr$^{-1}$ would lead to their current disk masses\footnote{The current disk mass estimates rely on dust mass, and accounting for enhanced radial drift and dust loss from external PE \citep{Sellek2020} could imply a higher disk gas-to-dust ratio, further shortening the timescale to reach the current disk mass.} in $\sim$0.85~Myr and $\sim$1.25~Myr, respectively, much lower than their estimated ages leading to the ``proplyd lifetime problem".
This suggests that either their initial disk masses were higher or that the external PE rate in the past was much lower, or both. The latter is equally possible for J1558-2310 since $\delta$~Sco is currently moving towards it (see Figure \ref{fig:Direction_confirmation} and its discussion in Section \ref{sec:confirmation_proplyds}), suggesting lower impact in the past \citep[see e.g.,][for the importance of cluster dynamics]{Scally2005,Winter2019}. The same could also be true for J1610-1919 since $\beta$~Sco is moving slower than J1610-1919 (Figure \ref{fig:Direction_confirmation}).

We can also estimate the remaining disk lifetime under the assumption that these disks are dissipated solely by external PE at their current mass loss rates. This yields remaining lifetimes of 5.6, 0.5, and 0.25~Myr for J1606-1841, J1558-2310, and J1610-1919, respectively, indicating that external PE affects J1558-2310 and J1610-1919 significantly more than J1606-1841. J1558-2310 is also currently accreting at a rate comparable to its wind mass-loss rate \citep[$\sim$10$^{-9}$~M$_{\odot}$~yr$^{-1}$,][]{Fang2023}, and $\delta$~Sco appears to be moving toward the disk (Figure~\ref{fig:Direction_confirmation}), suggesting that its remaining lifetime could actually be significantly shorter than 0.5~Myr.

Finally, for a typical T~Tauri system with stellar mass between 0.5 and 1~M$_{\odot}$, the lifetime-averaged internal (accretion + outflows) mass loss rate falls between $\sim$10$^{-8}$ and $\sim$10$^{-9}$~M$_{\odot}$~yr$^{-1}$ \citep[e.g.,][]{Hartmann1998,Manara2023,Pascucci2023}. Given the observed external PE mass loss rate of $\sim$10$^{-9}$~M$_{\odot}$~yr$^{-1}$ and the fact that external PE is a transient process \citep[e.g.,][]{Coleman2025b,Coleman2025a}, lasting up to a few Myr, the internal processes are more likely to dominate the dispersal and hence the lifetime of J1606-1841 (M$_* \sim$ 0.6~M$_{\odot}$). On the other hand, for a 0.1~M$_{\odot}$ star, the lifetime-averaged internal mass loss rate likely drops below 10$^{-9}$~M$_{\odot}$~yr$^{-1}$ \citep[also supported by the maximum observed accretion rate of 10$^{-8}$~M$_{\odot}$~yr$^{-1}$ for stellar masses between 0.1 and 0.2~M$_{\odot}$ in the nearby regions,][]{Manara2023}, making average external PE-driven mass loss comparable to the internal processes. Hence, the disks around low-mass stars seem to be significantly more impacted in low-FUV environments, to an extent where external PE can compete and likely even dominate over internal dispersal mechanisms. This is in line with theoretical expectations that smaller gravitational radii ($R_{grav}=GM_*/c_s^2$) around lower-mass stars can expose a larger fraction of the disk to external radiation \citep{Coleman2022,Haworth2023}, given that the gravitational radius has a steeper dependence on the stellar mass than the disk size ($R_{CO} \propto M_{\star}^{0.64}$, see Section \ref{sec:comparison_theoretical_pred_allsources}). This interpretation is also consistent with the recent result from \cite{Ping2026} that found a larger drop in the fraction of surviving disks around lower-mass stars in Upper Sco due to external PE, based on population synthesis.
 
\subsection{Comparison of theoretical and observed wind mass loss rates: Why only 3 molecular proplyds?} \label{sec:discuss_fried_comparison}

In Figure~\ref{fig:FRIED_4d_comparison}, we compared the theoretically predicted wind mass loss rates from \textsc{FRIEDv2} (details in Section~\ref{sec:comparison_theoretical_pred_allsources}) with several star and disk parameters for the 28 disks in the primary sample of the Upper Sco region observed with MIRI. While the two observed molecular proplyds (excluding J1610-1919, for which theoretical mass-loss rate using \textsc{FRIEDv2} cannot be estimated, see Section \ref{sec:comparison_theoretical_pred_allsources}) have predicted mass-loss rates in the top four, consistent with their detections, the question remains as to why other disks with similarly high predicted rates do not show the expected observational signatures of externally-driven PE winds. This is not attributable to per-pixel SNR, as other disks have comparable or higher SNR \citep[see][]{Bajaj2026}. Nor is it a matter of surface brightness sensitivity: other disks should in fact have better detectability for a similar H$_2$ line flux, since J1606-1841 has one of the lowest incident FUV fluxes ($\sim10~G_0$; see Table~\ref{tab:properties}). For disks experiencing higher FUV fluxes, the molecular dissociation front will move inward, closer to the disk, concentrating the line flux over fewer pixels than in J1606-1841.

\begin{figure}
    \centering
    \includegraphics[width=\linewidth]{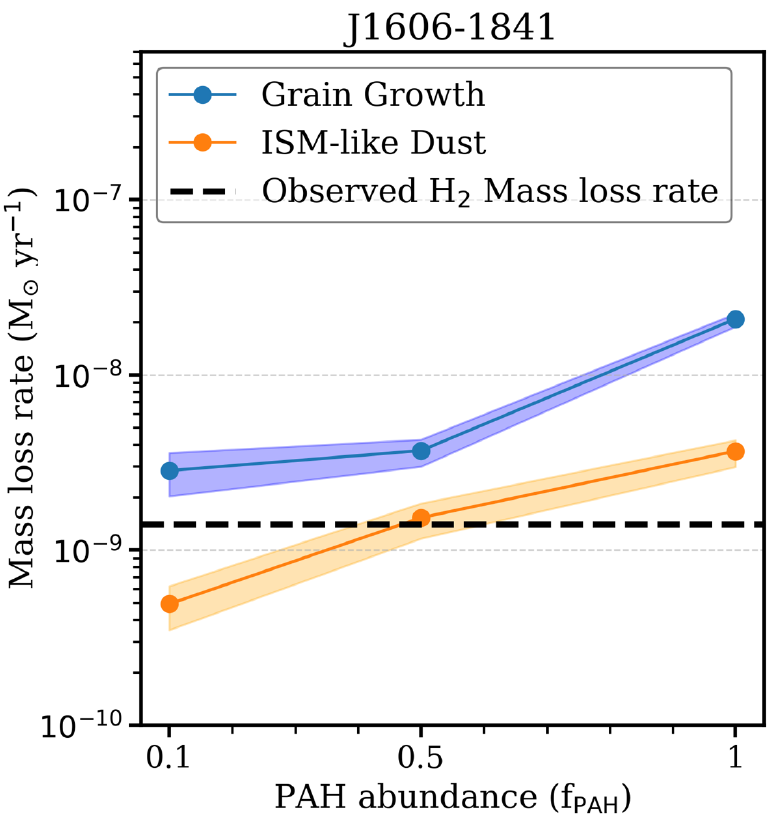}
    \caption{\textsc{FRIEDv2} predicted external PE wind mass-loss rates for J1606-1841 (the only one with observationally constrained gas disk size) with varying PAH abundances and dust grain/UV cross-sections (see Section \ref{sec:comparison_theoretical_pred_allsources} for details). The uncertainties correspond to the 16th and 84th percentile values of the incident FUV flux on the disk according to \cite{Anania2025a}. The horizontal black dashed line highlights the observed H$_2$ wind mass-loss rate (see Section \ref{sec:h2_mass_loss_calculation} for details).}
    \label{fig:PAH_abundance_massloss}
\end{figure}

One effect the \textsc{FRIEDv2} model grid does not account for, but which is likely important in weak FUV fields, is the relative 3D geometry of the disk and the nearby O/B star. \cite{Adams2004} analytically showed that the ratio of wind mass loss from the disk surface to that from the disk outer edge scales as the square root of the ratio of disk radius to the gravitational radius. In weak FUV fields, the sound speed decreases and the gravitational radius increases, so the fractional mass contribution from the disk outer edge grows. This was later corroborated by 2D external PE models \citep[assuming isotropic radiation field,][]{Haworth2019}, which found that already at 1000~G$_0$ the majority of mass loss originates from the disk outer edge, with only $\sim$20\% from the disk surface \citep[also see][]{Winter2022}. In such weak FUV fields, the disk will therefore lose more mass when its outer edge directly faces the irradiating star than when its surface does. Observationally, testing this is challenging due to two unknowns: (a) the accurate distance vector to the O/B star relative to the disk, i.e., whether the O/B star lies between the observer and the disk or beyond both, and (b) whether the disk is inclined toward or away from the observer. While both quantities can in principle be observationally determined, they are currently unavailable for a large fraction of our sample. Preliminarily, given the relatively high inclinations of J1606-1841 ($\sim$80$^\circ$, Table \ref{tab:properties}) and J1558-2310 ($\sim$55$^\circ$, Table \ref{tab:properties}), their disk position angles are roughly aligned with the dissociation front (see Figure \ref{fig:Direction_confirmation}), suggesting that indeed, the edges of their disks are likely facing the irradiating star.

J1610-1919 presents a test case. Although $\nu$~Sco and $\omega$~Sco have comparable FUV luminosities, $\nu$~Sco is estimated to be a factor of $\sim$2.5 closer to J1610–1919 than $\omega$~Sco in both three-dimensional and projected separations (see Table \ref{tab:properties}). Nevertheless, the dissociation front is directed toward the weaker (farther) irradiating source, $\omega$~Sco. We hypothesize that this is likely because the disk edge is oriented toward $\omega$~Sco, whereas the disk surface faces $\nu$~Sco, allowing irradiation from $\omega$~Sco to drive photoevaporation more effectively. Because the disk around J1610–1919 remains unresolved in existing ALMA observations, its inclination and position angle are currently unconstrained \citep{vanderplas2016}; however, future ALMA observations can test this hypothesis.

Finally, the mass-loss rates shown in Figure~\ref{fig:FRIED_4d_comparison} were computed for the fiducial \textsc{FRIEDv2} parameters, i.e., an ISM-like PAH-to-dust ratio ($f_{\rm PAH} = 1$) and the ``grain growth'' UV cross section in the wind ($\sigma_{\rm FUV} = 4.5\times10^{-23}$~cm$^2$, corresponding to a wind dust-to-gas mass ratio of $10^{-4}$). Varying these parameters can change the mass-loss rates by an order of magnitude \citep{Haworth2023}. This is illustrated for J1606-1841 in Figure~\ref{fig:PAH_abundance_massloss}, where the fiducial parameters yield the highest theoretical rates. Disks that lack clear signatures of external PE despite high predicted mass-loss rates may therefore have a larger UV cross section in the wind (i.e., a higher dust-to-gas mass ratio) and/or a lower PAH abundance than assumed. For J1606-1841, the observed H$_2$ wind mass-loss rate agrees well with the \textsc{FRIEDv2} parameter space. The agreement is best for ISM-like grains (wind dust-to-gas mass ratio of $10^{-2}$) with $f_{\rm PAH} \sim 0.5$ and/or for larger grains with a very low PAH abundance.
Such parameters, if extended to the rest of the disk population, would be expected to lower their mass-loss rates, meaning that fewer disks would be expected to have detectable mass-loss rates. Beyond this, a comparison of the \textsc{FRIEDv2} rates with a large sample of (as-yet-unidentified) low-FUV proplyds is needed to determine the reliability of its predicted rates.

\section{Summary and Conclusion} \label{sec:summary_conclusion}

In this study, we analyzed MIRI MRS observations of all Upper Sco disks with publicly available archival data to search for signatures of external photoevaporation. Of the 33 disks analyzed, we found three to exhibit cometary (proplyd-like) morphology in rotational H$_2$ emission lines \citep[two were recently reported in][]{Bajaj2026}. By comparing the orientation of the bright front with the stellar proper motion and with the positions of nearby massive stars, we found that the extended H$_2$ emission indeed traces externally driven photoevaporative winds in all three cases. We confirmed this further through pixel-by-pixel H$_2$ rotational diagram fits, which reveal elevated temperatures at the dissociation front. These fits also allowed us to estimate the wind mass-loss rates, which we then compared with theoretical predictions from the \textsc{FRIEDv2} grid of models \citep{Haworth2023}. Our major results are as follows:

\begin{enumerate}
    \item We found the H$_2$ dissociation fronts of J1606-1841 ($\sim$13~G$_0$), J1558-2310 ($\sim$45~G$_0$), and J1610-1919 ($\sim$33~G$_0$) to be pointed towards $\omega$~Sco (B1~V), $\delta$~Sco (B0.3~IV and B1-3~V binary) and $\omega$~Sco, respectively, making them the first direct detections of external photoevaporation in $<100$~G$_0$ environments. We did not detect an ionization front toward either disk in the atomic and ionized lines traced by MIRI. The I-front could lie beyond MIRI's FOV, or the flow could be entirely FUV-driven and lack a detectable I-front. In any case, these observations \citep[along with the results of][]{Booth2026} establish H$_2$ as the key observable tracer of external photoevaporation in weak UV environments ($<1000$~G$_0$).

    \item We analyzed H$_2$ emission toward 30 additional Upper Sco disks with archival MIRI observations and found no extended cometary morphology, even though theoretical models predict comparable (or higher) photoevaporation rates for some of them relative to J1606-1841. These non-detections suggest that current models omit an important factor: the relative 3D geometry of the disk and the external ionizing source. In weak environments, a massive star irradiating the edge of a disk should drive significantly more mass loss than one irradiating its face \citep{Adams2004}. The high inclinations of J1558-2310 and J1606-1841, combined with the close alignment between their disk position angles and the direction of the dissociation fronts (see Figure \ref{fig:Direction_confirmation}), support this interpretation.

    \item From the pixel-by-pixel H$_2$ rotational diagram fits, we derived temperatures of 500--700~K at the dissociation front with column densities of $\sim$10$^{19}$~cm$^{-2}$, corresponding to wind mass-loss rates of $\sim$10$^{-9}$~M$_{\odot}$~yr$^{-1}$. Given their current disk masses, this external PE wind could disperse the disk of J1610-1919 and J1558-2310 entirely within the next 0.25-0.5~Myr, whereas that of J1606-1841 could survive for $>5$~Myr, impacting the lifetime of the disk around lower-mass stars ($\sim$0.06 and 0.14~M$_{\odot}$) significantly more than that around a higher-mass star ($\sim$0.56~M$_{\odot}$). \cite{Ping2026} also reached the same conclusion using population synthesis modeling. For the low-mass disks, we find that the mass-loss rates driven by external PE can compete with those due to the internal processes of accretion and disk winds/jets.
    
\end{enumerate}

We identify several trends that warrant further testing. First, proplyds in fields below 1000~G$_0$ exhibit extended H$_2$ emission tracing a dissociation front offset from the disk, whereas above 1000~G$_0$ the emission, when detected, is compact. Second, in such weak-field environments, the lifetimes and dispersal of disks around low-mass stars ($\sim$0.1~M$_{\odot}$) are affected far more strongly than those around higher-mass stars ($>0.5$~M$_{\odot}$). Third, in these weak fields, the 3D geometry of the massive star relative to the disk likely plays an important role, in addition to the relative disk size for a given stellar mass. 
Confirming these trends will require assembling a larger sample of molecular proplyds spanning a broad range of UV field strengths, stellar masses, and viewing geometries, together with models that explicitly incorporate the relative 3D geometry of the disk and the external ionizing source.

Overall, the direct detection of externally driven photoevaporative winds toward nearby low-FUV disks opens a new window into the long-anticipated impact of such winds in the best-studied disk population. Their role in Upper Sco is particularly important, as this region is the only evolved counterpart of young star-forming regions that has so far been used to benchmark disk evolution and lifetimes. The detection of three proplyds among 33 disks traces the present-day incidence of external PE in Upper Sco. This fraction likely represents a lower limit on its cumulative impact on disk evolution because its time dependence is unknown, and many disks that are compact today may have undergone substantial external photoevaporation in the past \citep{Anania2025b,Ping2026}.

\begin{acknowledgments}

We thank the anonymous referee for their constructive suggestions and prompt feedback, which improved the manuscript. We thank Thomas Haworth and Uma Gorti for insightful discussions. We also thank Suzan Edwards for reading the manuscript and providing helpful comments. 

This work is based on observations made with the NASA/ESA/CSA James Webb Space Telescope. The JWST data presented in this article were obtained from the Mikulski Archive for Space Telescopes (MAST) at the Space Telescope Science Institute (STScI). The specific observations analyzed can be accessed via \dataset[doi: 10.17909/9gkt-eh36]{https://doi.org/10.17909/9gkt-eh36}. This research has made use of the NASA/IPAC Infrared Science Archive, which is funded by NASA and operated by the California Institute of Technology. This material is based on work supported by NASA under Grant No. 80NSSC25K0301 issued through the NNH24ZDA001N-FINESST program. N.S.B. and I.P. acknowledge partial support from NASA/STScI GO grants JWST-GO-01621.001 and JWST-GO-02260.001. Any opinions, findings, conclusions, or recommendations expressed in this article are those of the author(s) and do not necessarily reflect the views of NASA. A.D.S.'s contribution to this research was funded in part by the French National Research Agency (ANR) under grant number ANR-25-ERCS-0017 and by the European Research Council grant 101019751 MOLDISK. 
R.A. acknowledges funding by the European Union through the E-BEANS ERC project (grant number 100117693).
J.S.K. acknowledges NASA/STScI grant JWST-GO-06130 and NASA’s Nexus for Exoplanet System Science (NExSS) research coordination network sponsored by NASA’s Science Mission Directorate and project “Alien Earths” funded under Agreement No. 80NSSC21K0593.
The authors acknowledge use of GitHub Copilot through the Visual Studio Code extension to assist with debugging and optimizing Python data analysis, and Claude Opus 5.5 to assist with grammar refinement. All AI-assisted outputs were reviewed, verified, and edited by the authors.

\end{acknowledgments}

\facilities{James Webb Space Telescope, IRSA, WISE}

\software{Astropy \citep{astropy:2013,astropy:2018,astropy:2022}, IFU-Analyzer \citep{ifuanalyzer}, JWST \citep{Bushouse2026}, LMFIT \citep{lmfit}, Matplotlib \citep{Hunter:2007}, NumPy \citep{harris2020array}, Pandas \citep{pandas}, Photutils \citep{photutils}, Scikit-learn \citep{scikit-learn}, SciPy \citep{2020SciPy-NMeth}}

\appendix

\section{WISE maps of B stars} \label{sec:bow_shock_WISE}

\begin{figure*}
    \centering
    \includegraphics[width=\linewidth]{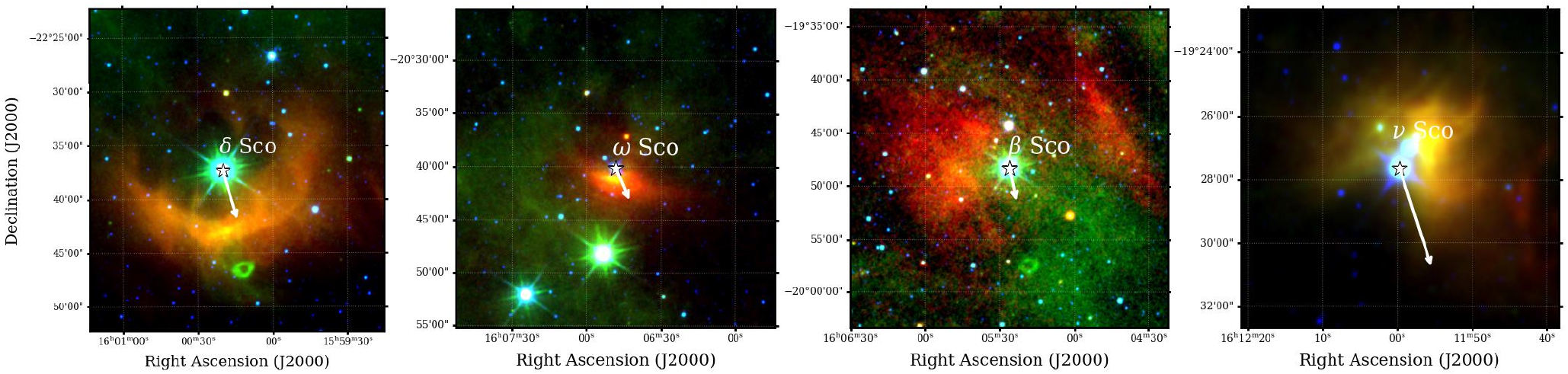}
    \caption{WISE W1, W3, and W4 three color images. The tip of the white arrow marks their predicted positions at J10000 using their current proper motions. $\delta$ Sco's bow-shock was first identified in \cite{Peri2012}.}
    \label{fig:bow_shock_WISE}
\end{figure*}

Figure~\ref{fig:bow_shock_WISE} shows Wide-field Infrared Survey Explorer (WISE) three-color images of $\delta$~Sco (the driving source of J1558-2310), $\omega$~Sco (the driving source of J1606-1841 and J1610-1919), $\beta$~Sco (dominant FUV source for J1606-1841), and $\nu$~Sco (dominant FUV source for J1610-1919). The images combine the W1 ($\sim$3.4~\micron{}), W3 ($\sim$12~\micron{}), and W4 ($\sim$22~\micron{}) bands, with data retrieved from the NASA/IPAC Infrared Science Archive (IRSA). The proper motion vectors are derived from data in \cite{Anania2025a}. The bow-shock towards $\delta$~Sco was previously identified by \cite{Peri2012}, whereas the possible bow-shock towards $\omega$~Sco is reported here for the first time.

\section{H$_2$ maps of all 33 MIRI-observed Upper Sco Disks} \label{sec:h2_line_maps}

The H$_2$ $\nu$0-0 S(1), S(3), and S(5) pixel-by-pixel continuum-subtracted line intensity maps of all 33 Upper Sco disks are shown in Figure \ref{fig:H2_maps_all_uppsco}. Proplyd-like emission morphology is seen only towards J1606-1841, J1558-2310 and J1610-1919. A few other noteworthy emission morphologies include an extension in the disk direction towards J16052157-1821412 and J16075796-2040087 to distances $\gtrsim$300~au. J15555600-2045187 shows a similar morphology; however, its disk position angle is unknown. A clear conical wind is seen towards J16082324-1930009 \citep[previously reported in][]{Bajaj2026}. H$_2$ emission from both binary components is observed towards J16153456-2242421 and AS205. Finally, 4 out of 5 brown dwarf disks show extended H$_2$ emission in at least one of the lines, with J1555-2045 tracing the disk or a wind depending on its position angle, J1559-2338 surrounded by diffuse limb-like S(1) emission, J1610-1919 tracing externally driven PE wind, and J1606-2056 showing strong S(3) emission centered on the star/disk system.

\begin{figure*}
    \centering
    \includegraphics[width=\linewidth]{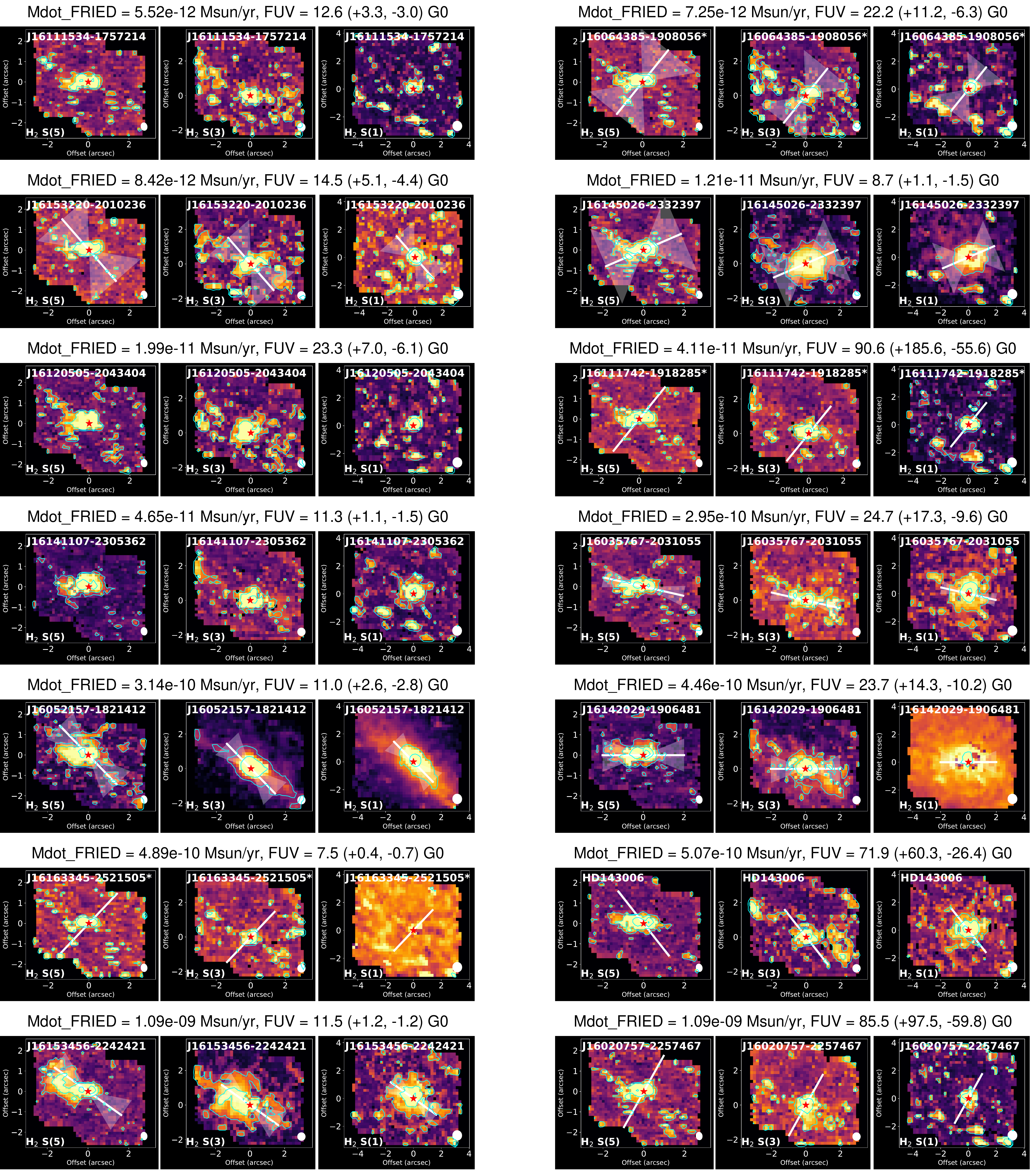}
    \caption{H$_2$~$\nu$0-0 S(5), S(3) and S(1) spaxel-by-spaxel continuum-subtracted and intensity-integrated line maps for all "Upper Sco" disks observed with MIRI through Cycle 3. All maps are displayed using an asinh normalization with a maximum percentile cut of 97.5. For maps with peak values below 35$\times \sigma$, a linear normalization with a maximum percentile cut of 97 is used instead. The cyan contours overlaid are 3$\sigma$, 15$\sigma$, 90$\sigma$, and 300$\sigma$, where $\sigma$ is the standard deviation of the background calculated iteratively using the \texttt{astropy} tool \texttt{sigma\_clipped\_stats}. The red star is placed at the centroid of the continuum near the line wavelength, and the white line passing through it highlights the disk position angle as calculated using spatially resolved ALMA continuum observations. The FWHM of the PSF at the line wavelength is shown at the bottom right in each map. The FUV flux at the source due to the nearby B stars and the FRIED-estimated mass loss rates for each source are shown above their H$_2$ maps. When the FRIED estimates are unavailable (for the five very-low-mass/brown-dwarf disks, Section \ref{sec:comparison_theoretical_pred_allsources}), the stellar masses are given instead.}
    \label{fig:H2_maps_all_uppsco}
\end{figure*}

\begin{figure*}
    \ContinuedFloat
    \centering
    \includegraphics[width=\linewidth]{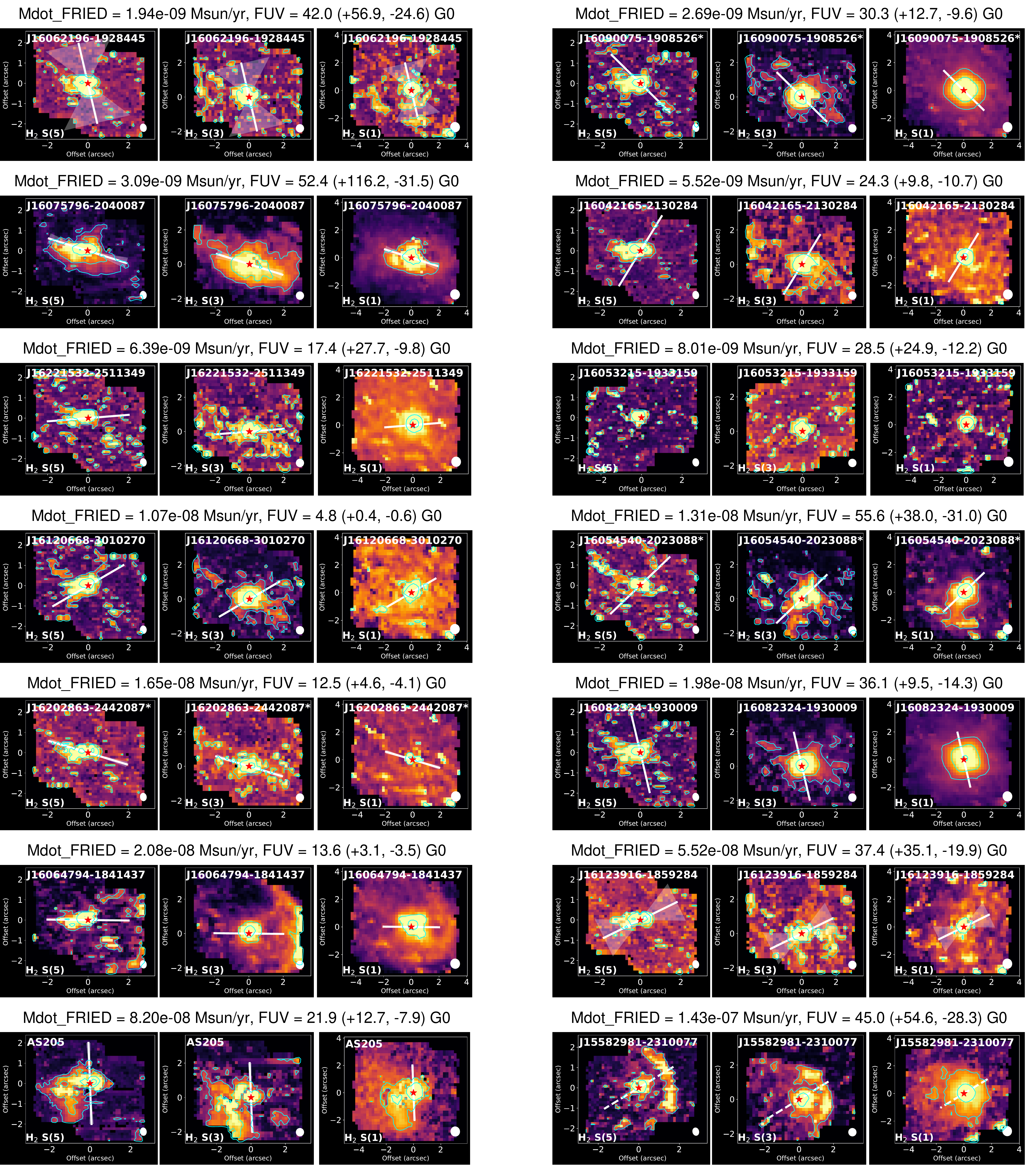}
    \caption{Continued}
\end{figure*}

\begin{figure*}
    \ContinuedFloat
    \centering
    \includegraphics[width=\linewidth]{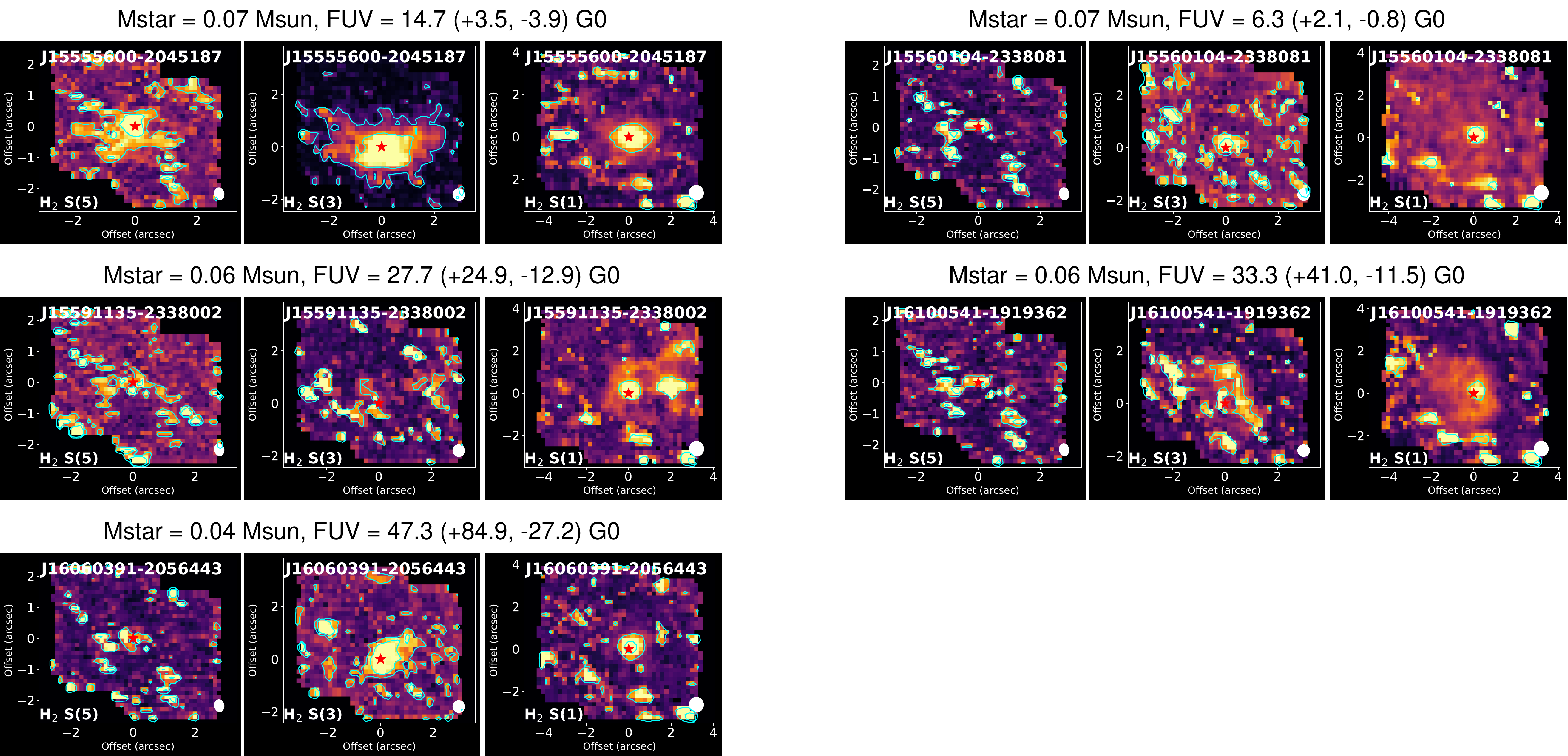}
    \caption{Continued}
\end{figure*}

\section{H$_2$ rotational diagram examples and mass estimation regions} \label{sec:example_rotational_diag}

\begin{figure*}
    \centering
    \includegraphics[width=\linewidth]{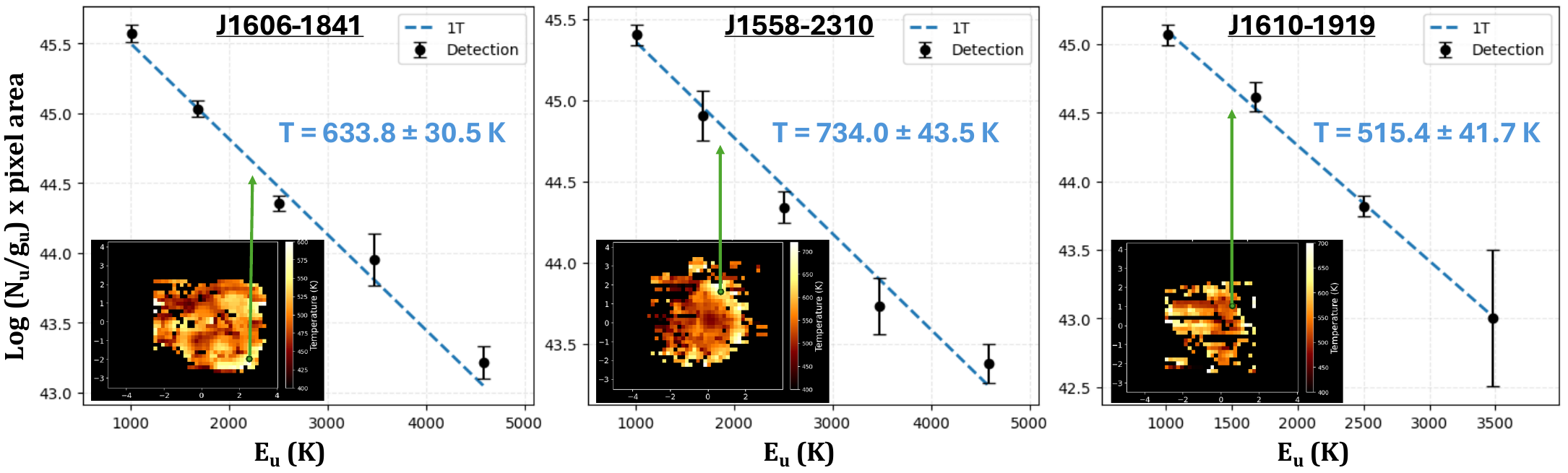}
    \caption{For each source, an example rotational diagram is shown, constructed at the pixel indicated by the green arrow. The y-axis represents the total number of molecules (i.e., column density x pixel area) populating the corresponding energy level. The best-fit temperature is indicated in each panel.}
    \label{fig:example_rotational_diag}
\end{figure*}

\begin{figure*}
    \centering
    \includegraphics[width=\linewidth]{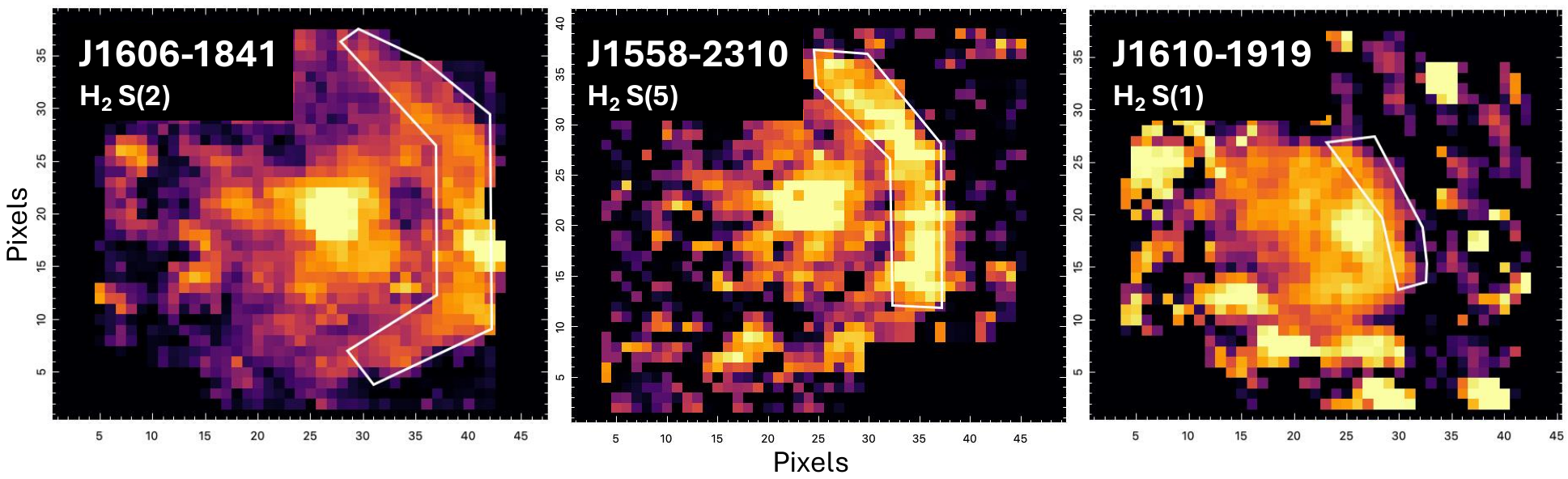}
    \caption{H$_2$ line intensity maps highlighting the regions where the H$_2$ masses were summed, as described in Section \ref{sec:h2_mass_loss_calculation}. The pixel sizes are 0.2'', 0.13'', and 0.2'', respectively.}
    \label{fig:mass_region_highlighted}
\end{figure*}

As described in Section~\ref{sec:h2_mass_loss_calculation}, we derived temperature and column density maps from pixel-by-pixel rotational-diagram fits. Figure~\ref{fig:example_rotational_diag} shows an example fit at the dissociation front of each source. For each fit, we included only the H$_2$ $\nu$0-0 S(1)-S(5) lines because the S(6)-S(8) lines are not detected at the dissociation fronts. A fit was performed where at least three lines were detected at $>3\sigma$. Additionally, Figure \ref{fig:mass_region_highlighted} highlights the regions where the H$_2$ masses were summed to obtain the wind mass loss rates. The masses derived from these regions represent lower limits on the total wind mass and hence, mass loss rates (see Section \ref{sec:h2_mass_loss_calculation}).

\bibliography{citations}{}

\begin{thebibliography}{}
\expandafter\ifx\csname natexlab\endcsname\relax\def\natexlab#1{#1}\fi
\providecommand{\url}[1]{\href{#1}{#1}}
\providecommand{\dodoi}[1]{doi:~\href{http://doi.org/#1}{\nolinkurl{#1}}}
\providecommand{\doeprint}[1]{\href{http://ascl.net/#1}{\nolinkurl{http://ascl.net/#1}}}
\providecommand{\doarXiv}[1]{\href{https://arxiv.org/abs/#1}{\nolinkurl{https://arxiv.org/abs/#1}}}

\bibitem[{{Adams} {et~al.}(2004){Adams}, {Hollenbach}, {Laughlin}, \& {Gorti}}]{Adams2004}
{Adams}, F.~C., {Hollenbach}, D., {Laughlin}, G., \& {Gorti}, U. 2004, \apj, 611, 360, \dodoi{10.1086/421989}

\bibitem[{{Amiot} {et~al.}(2026){Amiot}, {Bern{\'e}}, {Schroetter}, {Robberto}, {Haworth}, {Boersma}, {Dartois}, {Fuente}, {Goicoechea}, {Habart}, {McCaughrean}, {Onaka}, \& {Peeters}}]{Amiot2026}
{Amiot}, P., {Bern{\'e}}, O., {Schroetter}, I., {et~al.} 2026, \aap, 708, A111, \dodoi{10.1051/0004-6361/202558551}

\bibitem[{{Anania} {et~al.}(2025{\natexlab{a}}){Anania}, {Winter}, {Rosotti}, {Vioque}, {Zari}, {Pantaleoni Gonz{\'a}lez}, \& {Testi}}]{Anania2025a}
{Anania}, R., {Winter}, A.~J., {Rosotti}, G., {et~al.} 2025{\natexlab{a}}, \aap, 695, A74, \dodoi{10.1051/0004-6361/202453011}

\bibitem[{{Anania} {et~al.}(2025{\natexlab{b}}){Anania}, {Rosotti}, {G{\'a}rate}, {Pinilla}, {Vioque}, {Trapman}, {Carpenter}, {Zhang}, {Pascucci}, {Cieza}, {Sierra}, {Kurtovic}, {Miley}, {P{\'e}rez}, {Tabone}, {Hogerheijde}, {Deng}, {Agurto-Gangas}, {Ruiz-Rodriguez}, {Gonz{\'a}lez-Ruilova}, \& {TorresVillanueva}}]{Anania2025b}
{Anania}, R., {Rosotti}, G.~P., {G{\'a}rate}, M., {et~al.} 2025{\natexlab{b}}, \apj, 989, 8, \dodoi{10.3847/1538-4357/adb587}

\bibitem[{{Ansdell} {et~al.}(2017){Ansdell}, {Williams}, {Manara}, {Miotello}, {Facchini}, {van der Marel}, {Testi}, \& {van Dishoeck}}]{Ansdell2017}
{Ansdell}, M., {Williams}, J.~P., {Manara}, C.~F., {et~al.} 2017, \aj, 153, 240, \dodoi{10.3847/1538-3881/aa69c0}

\bibitem[{{Ansdell} {et~al.}(2016){Ansdell}, {Williams}, {van der Marel}, {Carpenter}, {Guidi}, {Hogerheijde}, {Mathews}, {Manara}, {Miotello}, {Natta}, {Oliveira}, {Tazzari}, {Testi}, {van Dishoeck}, \& {van Terwisga}}]{Ansdell2016}
{Ansdell}, M., {Williams}, J.~P., {van der Marel}, N., {et~al.} 2016, \apj, 828, 46, \dodoi{10.3847/0004-637X/828/1/46}

\bibitem[{{Ansdell} {et~al.}(2020){Ansdell}, {Gaidos}, {Hedges}, {Tazzari}, {Kraus}, {Wyatt}, {Kennedy}, {Williams}, {Mann}, {Angelo}, {D{\^u}chene}, {Mamajek}, {Carpenter}, {Esplin}, \& {Rizzuto}}]{Ansdell2020}
{Ansdell}, M., {Gaidos}, E., {Hedges}, C., {et~al.} 2020, \mnras, 492, 572, \dodoi{10.1093/mnras/stz3361}

\bibitem[{{Aru} {et~al.}(2024){Aru}, {Mauc{\'o}}, {Manara}, {Haworth}, {Facchini}, {McLeod}, {Miotello}, {Petr-Gotzens}, {Robberto}, {Rosotti}, {Vicente}, {Winter}, \& {Ansdell}}]{Aru2024}
{Aru}, M.-L., {Mauc{\'o}}, K., {Manara}, C.~F., {et~al.} 2024, \aap, 687, A93, \dodoi{10.1051/0004-6361/202349004}

\bibitem[{{Arulanantham} {et~al.}(2025){Arulanantham}, {Salyk}, {Pontoppidan}, {Banzatti}, {Zhang}, {{\"O}berg}, {Long}, {Carr}, {Najita}, {Pascucci}, {Colmenares}, {Xie}, {Huang}, {Green}, {Andrews}, {Blake}, {Bergin}, {Pinilla}, {Vioque}, {Dahl}, {Raul}, {Krijt}, \& {The Jdiscs Collaboration}}]{Arulanantham2025}
{Arulanantham}, N., {Salyk}, C., {Pontoppidan}, K., {et~al.} 2025, \aj, 170, 67, \dodoi{10.3847/1538-3881/addd01}

\bibitem[{{Astropy Collaboration} {et~al.}(2013){Astropy Collaboration}, {Robitaille}, {Tollerud}, {Greenfield}, {Droettboom}, {Bray}, {Aldcroft}, {Davis}, {Ginsburg}, {Price-Whelan}, {Kerzendorf}, {Conley}, {Crighton}, {Barbary}, {Muna}, {Ferguson}, {Grollier}, {Parikh}, {Nair}, {Unther}, {Deil}, {Woillez}, {Conseil}, {Kramer}, {Turner}, {Singer}, {Fox}, {Weaver}, {Zabalza}, {Edwards}, {Azalee Bostroem}, {Burke}, {Casey}, {Crawford}, {Dencheva}, {Ely}, {Jenness}, {Labrie}, {Lim}, {Pierfederici}, {Pontzen}, {Ptak}, {Refsdal}, {Servillat}, \& {Streicher}}]{astropy:2013}
{Astropy Collaboration}, {Robitaille}, T.~P., {Tollerud}, E.~J., {et~al.} 2013, \aap, 558, A33, \dodoi{10.1051/0004-6361/201322068}

\bibitem[{{Astropy Collaboration} {et~al.}(2018){Astropy Collaboration}, {Price-Whelan}, {Sip{\H{o}}cz}, {G{\"u}nther}, {Lim}, {Crawford}, {Conseil}, {Shupe}, {Craig}, {Dencheva}, {Ginsburg}, {Vand erPlas}, {Bradley}, {P{\'e}rez-Su{\'a}rez}, {de Val-Borro}, {Aldcroft}, {Cruz}, {Robitaille}, {Tollerud}, {Ardelean}, {Babej}, {Bach}, {Bachetti}, {Bakanov}, {Bamford}, {Barentsen}, {Barmby}, {Baumbach}, {Berry}, {Biscani}, {Boquien}, {Bostroem}, {Bouma}, {Brammer}, {Bray}, {Breytenbach}, {Buddelmeijer}, {Burke}, {Calderone}, {Cano Rodr{\'\i}guez}, {Cara}, {Cardoso}, {Cheedella}, {Copin}, {Corrales}, {Crichton}, {D'Avella}, {Deil}, {Depagne}, {Dietrich}, {Donath}, {Droettboom}, {Earl}, {Erben}, {Fabbro}, {Ferreira}, {Finethy}, {Fox}, {Garrison}, {Gibbons}, {Goldstein}, {Gommers}, {Greco}, {Greenfield}, {Groener}, {Grollier}, {Hagen}, {Hirst}, {Homeier}, {Horton}, {Hosseinzadeh}, {Hu}, {Hunkeler}, {Ivezi{\'c}}, {Jain}, {Jenness}, {Kanarek}, {Kendrew}, {Kern}, {Kerzendorf}, {Khvalko}, {King}, {Kirkby}, {Kulkarni},
  {Kumar}, {Lee}, {Lenz}, {Littlefair}, {Ma}, {Macleod}, {Mastropietro}, {McCully}, {Montagnac}, {Morris}, {Mueller}, {Mumford}, {Muna}, {Murphy}, {Nelson}, {Nguyen}, {Ninan}, {N{\"o}the}, {Ogaz}, {Oh}, {Parejko}, {Parley}, {Pascual}, {Patil}, {Patil}, {Plunkett}, {Prochaska}, {Rastogi}, {Reddy Janga}, {Sabater}, {Sakurikar}, {Seifert}, {Sherbert}, {Sherwood-Taylor}, {Shih}, {Sick}, {Silbiger}, {Singanamalla}, {Singer}, {Sladen}, {Sooley}, {Sornarajah}, {Streicher}, {Teuben}, {Thomas}, {Tremblay}, {Turner}, {Terr{\'o}n}, {van Kerkwijk}, {de la Vega}, {Watkins}, {Weaver}, {Whitmore}, {Woillez}, {Zabalza}, \& {Astropy Contributors}}]{astropy:2018}
{Astropy Collaboration}, {Price-Whelan}, A.~M., {Sip{\H{o}}cz}, B.~M., {et~al.} 2018, \aj, 156, 123, \dodoi{10.3847/1538-3881/aabc4f}

\bibitem[{{Astropy Collaboration} {et~al.}(2022){Astropy Collaboration}, {Price-Whelan}, {Lim}, {Earl}, {Starkman}, {Bradley}, {Shupe}, {Patil}, {Corrales}, {Brasseur}, {N{"o}the}, {Donath}, {Tollerud}, {Morris}, {Ginsburg}, {Vaher}, {Weaver}, {Tocknell}, {Jamieson}, {van Kerkwijk}, {Robitaille}, {Merry}, {Bachetti}, {G{"u}nther}, {Aldcroft}, {Alvarado-Montes}, {Archibald}, {B{'o}di}, {Bapat}, {Barentsen}, {Baz{'a}n}, {Biswas}, {Boquien}, {Burke}, {Cara}, {Cara}, {Conroy}, {Conseil}, {Craig}, {Cross}, {Cruz}, {D'Eugenio}, {Dencheva}, {Devillepoix}, {Dietrich}, {Eigenbrot}, {Erben}, {Ferreira}, {Foreman-Mackey}, {Fox}, {Freij}, {Garg}, {Geda}, {Glattly}, {Gondhalekar}, {Gordon}, {Grant}, {Greenfield}, {Groener}, {Guest}, {Gurovich}, {Handberg}, {Hart}, {Hatfield-Dodds}, {Homeier}, {Hosseinzadeh}, {Jenness}, {Jones}, {Joseph}, {Kalmbach}, {Karamehmetoglu}, {Ka{l}uszy{'n}ski}, {Kelley}, {Kern}, {Kerzendorf}, {Koch}, {Kulumani}, {Lee}, {Ly}, {Ma}, {MacBride}, {Maljaars}, {Muna}, {Murphy}, {Norman}, {O'Steen},
  {Oman}, {Pacifici}, {Pascual}, {Pascual-Granado}, {Patil}, {Perren}, {Pickering}, {Rastogi}, {Roulston}, {Ryan}, {Rykoff}, {Sabater}, {Sakurikar}, {Salgado}, {Sanghi}, {Saunders}, {Savchenko}, {Schwardt}, {Seifert-Eckert}, {Shih}, {Jain}, {Shukla}, {Sick}, {Simpson}, {Singanamalla}, {Singer}, {Singhal}, {Sinha}, {Sip{H{o}}cz}, {Spitler}, {Stansby}, {Streicher}, {{{S}}umak}, {Swinbank}, {Taranu}, {Tewary}, {Tremblay}, {Val-Borro}, {Van Kooten}, {Vasovi{'c}}, {Verma}, {de Miranda Cardoso}, {Williams}, {Wilson}, {Winkel}, {Wood-Vasey}, {Xue}, {Yoachim}, {Zhang}, {Zonca}, \& {Astropy Project Contributors}}]{astropy:2022}
{Astropy Collaboration}, {Price-Whelan}, A.~M., {Lim}, P.~L., {et~al.} 2022, \apj, 935, 167, \dodoi{10.3847/1538-4357/ac7c74}

\bibitem[{{Bajaj} {et~al.}(2025){Bajaj}, {Pascucci}, {Beck}, {Edwards}, {Cabrit}, {Najita}, {Schwarz}, {Semenov}, {Salyk}, {Gorti}, {Brittain}, {Krijt}, {Ruaud}, \& {Page}}]{Bajaj2025}
{Bajaj}, N.~S., {Pascucci}, I., {Beck}, T.~L., {et~al.} 2025, \aj, 169, 296, \dodoi{10.3847/1538-3881/adc73c}

\bibitem[{{Bajaj} {et~al.}(2026){Bajaj}, {Pascucci}, {Cabrit}, {Edwards}, {Cugno}, {Sellek}, {Najita}, {Zhang}, {Alexander}, {Herczeg}, {Gorti}, {Clark}, \& {Beck}}]{Bajaj2026}
{Bajaj}, N.~S., {Pascucci}, I., {Cabrit}, S., {et~al.} 2026, arXiv e-prints, arXiv:2607.21733, \dodoi{10.48550/arXiv.2607.21733}

\bibitem[{{Ballering} {et~al.}(2023){Ballering}, {Cleeves}, {Haworth}, {Bally}, {Eisner}, {Ginsburg}, {Boyden}, {Fang}, \& {Kim}}]{Ballering2023}
{Ballering}, N.~P., {Cleeves}, L.~I., {Haworth}, T.~J., {et~al.} 2023, \apj, 954, 127, \dodoi{10.3847/1538-4357/ace901}

\bibitem[{{Bally} {et~al.}(2000){Bally}, {O'Dell}, \& {McCaughrean}}]{Bally2000}
{Bally}, J., {O'Dell}, C.~R., \& {McCaughrean}, M.~J. 2000, \aj, 119, 2919, \dodoi{10.1086/301385}

\bibitem[{{Bally} {et~al.}(1998){Bally}, {Testi}, {Sargent}, \& {Carlstrom}}]{Bally1998b}
{Bally}, J., {Testi}, L., {Sargent}, A., \& {Carlstrom}, J. 1998, \aj, 116, 854, \dodoi{10.1086/300469}

\bibitem[{{Baraffe} {et~al.}(2015){Baraffe}, {Homeier}, {Allard}, \& {Chabrier}}]{Baraffe2015}
{Baraffe}, I., {Homeier}, D., {Allard}, F., \& {Chabrier}, G. 2015, \aap, 577, A42, \dodoi{10.1051/0004-6361/201425481}

\bibitem[{{Beckwith} {et~al.}(1990){Beckwith}, {Sargent}, {Chini}, \& {Guesten}}]{Beckwith1990}
{Beckwith}, S. V.~W., {Sargent}, A.~I., {Chini}, R.~S., \& {Guesten}, R. 1990, \aj, 99, 924, \dodoi{10.1086/115385}

\bibitem[{{Booth} {et~al.}(2026){Booth}, {Zhou}, {Kim}, {{\"O}berg}, {Calahan}, {Pontoppidan}, {Haworth}, {Pascucci}, {Boyden}, {Narang}, {Mauc{\'o}}, {Arabhavi}, \& {Ballering}}]{Booth2026}
{Booth}, A.~S., {Zhou}, Q., {Kim}, J.~S., {et~al.} 2026, arXiv e-prints, arXiv:2608.17226, \dodoi{10.48550/arXiv.2608.17226}

\bibitem[{{Boyden} \& {Eisner}(2020)}]{Boyden2020}
{Boyden}, R.~D., \& {Eisner}, J.~A. 2020, \apj, 894, 74, \dodoi{10.3847/1538-4357/ab86b7}

\bibitem[{{Boyden} \& {Eisner}(2024)}]{Boyden2024}
---. 2024, \apj, 967, 103, \dodoi{10.3847/1538-4357/ad3cd5}

\bibitem[{Bradley {et~al.}(2026)Bradley, Sipőcz, Robitaille, Tollerud, Vinícius, Deil, Barbary, Wilson, Busko, Donath, Günther, Cara, Lim, Meßlinger, Conseil, Droettboom, Bostroem, Bray, Bratholm, Burnett, Jamieson, Ginsburg, Taranu, Barentsen, Craig, Morris, Perrin, \& Rathi}]{photutils}
Bradley, L., Sipőcz, B.~M., Robitaille, T.~P., {et~al.} 2026, Photutils, 3.0.0,  Zenodo, \dodoi{10.5281/zenodo.19636730}

\bibitem[{{Bressan} {et~al.}(2012){Bressan}, {Marigo}, {Girardi}, {Salasnich}, {Dal Cero}, {Rubele}, \& {Nanni}}]{Bressan2012}
{Bressan}, A., {Marigo}, P., {Girardi}, L., {et~al.} 2012, \mnras, 427, 127, \dodoi{10.1111/j.1365-2966.2012.21948.x}

\bibitem[{{Bushouse} {et~al.}(2026){Bushouse}, {Eisenhamer}, {Dencheva}, {Davies}, {Greenfield}, {Morrison}, {Hodge}, {Simon}, {Grumm}, {Droettboom}, {Slavich}, {Sosey}, {Pauly}, {Miller}, {Jedrzejewski}, {Hack}, {Davis}, {Crawford}, {Law}, {Gordon}, {Regan}, {Cara}, {MacDonald}, {Bradley}, {Shanahan}, {Jamieson}, {Teodoro}, {Williams}, {Pena-Guerrero}, {Graham}, {Molter}, {Brandt}, {Hayes}, {Cooper}, {Clarke}, \& {Filippazzo}}]{Bushouse2026}
{Bushouse}, H., {Eisenhamer}, J., {Dencheva}, N., {et~al.} 2026, {JWST Calibration Pipeline}, 2.0.1,  Zenodo, \dodoi{10.5281/zenodo.20058613}

\bibitem[{{Carpenter} {et~al.}(2025){Carpenter}, {Esplin}, {Luhman}, {Mamajek}, \& {Andrews}}]{Carpenter2025}
{Carpenter}, J.~M., {Esplin}, T.~L., {Luhman}, K.~L., {Mamajek}, E.~E., \& {Andrews}, S.~M. 2025, \apj, 978, 117, \dodoi{10.3847/1538-4357/ad8ebc}

\bibitem[{{Chen} {et~al.}(2015){Chen}, {Bressan}, {Girardi}, {Marigo}, {Kong}, \& {Lanza}}]{Chen2015}
{Chen}, Y., {Bressan}, A., {Girardi}, L., {et~al.} 2015, \mnras, 452, 1068, \dodoi{10.1093/mnras/stv1281}

\bibitem[{{Clarke} \& {Owen}(2015)}]{Clarke2015}
{Clarke}, C.~J., \& {Owen}, J.~E. 2015, \mnras, 446, 2944, \dodoi{10.1093/mnras/stu2248}

\bibitem[{{Coleman} \& {Haworth}(2022)}]{Coleman2022}
{Coleman}, G. A.~L., \& {Haworth}, T.~J. 2022, \mnras, 514, 2315, \dodoi{10.1093/mnras/stac1513}

\bibitem[{{Coleman} {et~al.}(2025{\natexlab{a}}){Coleman}, {Haworth}, \& {Kim}}]{Coleman2025b}
{Coleman}, G. A.~L., {Haworth}, T.~J., \& {Kim}, J.~S. 2025{\natexlab{a}}, \mnras, 544, L70, \dodoi{10.1093/mnrasl/slaf100}

\bibitem[{{Coleman} {et~al.}(2025{\natexlab{b}}){Coleman}, {Kim}, {Haworth}, {Hartman}, \& {Kalish}}]{Coleman2025a}
{Coleman}, G. A.~L., {Kim}, J.~S., {Haworth}, T.~J., {Hartman}, P.~A., \& {Kalish}, T.~C. 2025{\natexlab{b}}, \mnras, 540, 1202, \dodoi{10.1093/mnras/staf789}

\bibitem[{{D{\'\i}az-Berr{\'\i}os} {et~al.}(2024){D{\'\i}az-Berr{\'\i}os}, {Guzm{\'a}n}, {Walsh}, {{\"O}berg}, {Cleeves}, {Artur de la Villarmois}, \& {Carpenter}}]{Diaz-Berrios2024}
{D{\'\i}az-Berr{\'\i}os}, J.~K., {Guzm{\'a}n}, V.~V., {Walsh}, C., {et~al.} 2024, \apj, 969, 165, \dodoi{10.3847/1538-4357/ad4603}

\bibitem[{{Eisner} {et~al.}(2008){Eisner}, {Plambeck}, {Carpenter}, {Corder}, {Qi}, \& {Wilner}}]{Eisner2008}
{Eisner}, J.~A., {Plambeck}, R.~L., {Carpenter}, J.~M., {et~al.} 2008, \apj, 683, 304, \dodoi{10.1086/588524}

\bibitem[{{Empey} {et~al.}(2026){Empey}, {Manara}, {Garcia Lopez}, {Natta}, {Claes}, {Zagaria}, {Alcal{\'a}}, {Anania}, {Beccari}, {Carpenter}, {Facchini}, {Fedele}, {Lodato}, {Mauco}, {Miotello}, {Nisini}, {Pascucci}, {Piscarreta}, {Rosotti}, {Scholz}, {Testi}, \& {Vioque}}]{Empey2026}
{Empey}, A., {Manara}, C.~F., {Garcia Lopez}, R., {et~al.} 2026, \aap, 712, A222, \dodoi{10.1051/0004-6361/202660084}

\bibitem[{{Facchini} {et~al.}(2016){Facchini}, {Clarke}, \& {Bisbas}}]{Facchini2016}
{Facchini}, S., {Clarke}, C.~J., \& {Bisbas}, T.~G. 2016, \mnras, 457, 3593, \dodoi{10.1093/mnras/stw240}

\bibitem[{{Fang} {et~al.}(2023){Fang}, {Pascucci}, {Edwards}, {Gorti}, {Hillenbrand}, \& {Carpenter}}]{Fang2023}
{Fang}, M., {Pascucci}, I., {Edwards}, S., {et~al.} 2023, \apj, 945, 112, \dodoi{10.3847/1538-4357/acb2c9}

\bibitem[{{Grant} {et~al.}(2023){Grant}, {Argyriou}, {Barrado Navascues}, {Christiaens}, {Gasman}, {Guedel}, {Henning}, {Kamp}, {Kanwar}, {Lagage}, \& et~al.}]{Grant2023}
{Grant}, S., {Argyriou}, I., {Barrado Navascues}, D., {et~al.} 2023, {Probing carbon chemistry and dust in the planet-forming zones of brown dwarf disks}, JWST Proposal. Cycle 2, ID. \#3886

\bibitem[{{Habing}(1968)}]{Habing1968}
{Habing}, H.~J. 1968, \bain, 19, 421

\bibitem[{Harris {et~al.}(2020)Harris, Millman, van~der Walt, Gommers, Virtanen, Cournapeau, Wieser, Taylor, Berg, Smith, Kern, Picus, Hoyer, van Kerkwijk, Brett, Haldane, del R{\'{i}}o, Wiebe, Peterson, G{\'{e}}rard-Marchant, Sheppard, Reddy, Weckesser, Abbasi, Gohlke, \& Oliphant}]{harris2020array}
Harris, C.~R., Millman, K.~J., van~der Walt, S.~J., {et~al.} 2020, Nature, 585, 357, \dodoi{10.1038/s41586-020-2649-2}

\bibitem[{{Hartmann} {et~al.}(1998){Hartmann}, {Calvet}, {Gullbring}, \& {D'Alessio}}]{Hartmann1998}
{Hartmann}, L., {Calvet}, N., {Gullbring}, E., \& {D'Alessio}, P. 1998, \apj, 495, 385, \dodoi{10.1086/305277}

\bibitem[{{Haworth} \& {Clarke}(2019)}]{Haworth2019}
{Haworth}, T.~J., \& {Clarke}, C.~J. 2019, \mnras, 485, 3895, \dodoi{10.1093/mnras/stz706}

\bibitem[{{Haworth} {et~al.}(2018){Haworth}, {Clarke}, {Rahman}, {Winter}, \& {Facchini}}]{Haworth2018}
{Haworth}, T.~J., {Clarke}, C.~J., {Rahman}, W., {Winter}, A.~J., \& {Facchini}, S. 2018, \mnras, 481, 452, \dodoi{10.1093/mnras/sty2323}

\bibitem[{{Haworth} {et~al.}(2023){Haworth}, {Coleman}, {Qiao}, {Sellek}, \& {Askari}}]{Haworth2023}
{Haworth}, T.~J., {Coleman}, G. A.~L., {Qiao}, L., {Sellek}, A.~D., \& {Askari}, K. 2023, \mnras, 526, 4315, \dodoi{10.1093/mnras/stad3054}

\bibitem[{{Haworth} {et~al.}(2017){Haworth}, {Facchini}, {Clarke}, \& {Cleeves}}]{Haworth2017}
{Haworth}, T.~J., {Facchini}, S., {Clarke}, C.~J., \& {Cleeves}, L.~I. 2017, \mnras, 468, L108, \dodoi{10.1093/mnrasl/slx037}

\bibitem[{{Haworth} {et~al.}(2021){Haworth}, {Kim}, {Winter}, {Hines}, {Clarke}, {Sellek}, {Ballabio}, \& {Stapelfeldt}}]{Haworth2021}
{Haworth}, T.~J., {Kim}, J.~S., {Winter}, A.~J., {et~al.} 2021, \mnras, 501, 3502, \dodoi{10.1093/mnras/staa3918}

\bibitem[{{Henney} \& {O'Dell}(1999)}]{Henney1999}
{Henney}, W.~J., \& {O'Dell}, C.~R. 1999, \aj, 118, 2350, \dodoi{10.1086/301087}

\bibitem[{{Henney} {et~al.}(2002){Henney}, {O'Dell}, {Meaburn}, {Garrington}, \& {Lopez}}]{Henney2002}
{Henney}, W.~J., {O'Dell}, C.~R., {Meaburn}, J., {Garrington}, S.~T., \& {Lopez}, J.~A. 2002, \apj, 566, 315, \dodoi{10.1086/338055}

\bibitem[{{Henning} {et~al.}(2024){Henning}, {Kamp}, {Samland}, {Arabhavi}, {Kanwar}, {van Dishoeck}, {G{\"u}del}, {Lagage}, {Waelkens}, {Abergel}, {Absil}, {Barrado}, {Boccaletti}, {Bouwman}, {Caratti o Garatti}, {Geers}, {Glauser}, {Lahuis}, {Mueller}, {Nehm{\'e}}, {Olofsson}, {Pantin}, {Ray}, {Scheithauer}, {Vandenbussche}, {Waters}, {Wright}, {Argyriou}, {Christiaens}, {Franceschi}, {Gasman}, {Grant}, {Guadarrama}, {Jang}, {Morales-Calder{\'o}n}, {Pawellek}, {Perotti}, {Rodgers-Lee}, {Schreiber}, {Schwarz}, {Tabone}, {Temmink}, {Vlasblom}, {Colina}, {Greve}, \& {{\"O}stlin}}]{Henning2024}
{Henning}, T., {Kamp}, I., {Samland}, M., {et~al.} 2024, \pasp, 136, 054302, \dodoi{10.1088/1538-3873/ad3455}

\bibitem[{{Henning} {et~al.}(2017){Henning}, {Abergel}, {Absil}, {Barrado Navascues}, {Boccaletti}, {Bouwman}, {Caratti o Garatti}, {Geers}, {Glauser}, {Guedel}, \& et~al.}]{Henning2017}
{Henning}, T.~K., {Abergel}, A., {Absil}, O., {et~al.} 2017, {MIRI EC Protoplanetary and Debris Disks Survey}, JWST Proposal. Cycle 1, ID. \#1282

\bibitem[{{Hildebrand}(1983)}]{Hildebrand1983}
{Hildebrand}, R.~H. 1983, \qjras, 24, 267

\bibitem[{Hunter(2007)}]{Hunter:2007}
Hunter, J.~D. 2007, Computing in Science \& Engineering, 9, 90, \dodoi{10.1109/MCSE.2007.55}

\bibitem[{{Johnstone} {et~al.}(1998){Johnstone}, {Hollenbach}, \& {Bally}}]{Johnstone1998}
{Johnstone}, D., {Hollenbach}, D., \& {Bally}, J. 1998, \apj, 499, 758, \dodoi{10.1086/305658}

\bibitem[{{Kim} {et~al.}(2016){Kim}, {Clarke}, {Fang}, \& {Facchini}}]{Kim2016}
{Kim}, J.~S., {Clarke}, C.~J., {Fang}, M., \& {Facchini}, S. 2016, \apjl, 826, L15, \dodoi{10.3847/2041-8205/826/1/L15}

\bibitem[{{Law} {et~al.}(2023){Law}, {E. Morrison}, {Argyriou}, {Patapis}, {{\'A}lvarez-M{\'a}rquez}, {Labiano}, \& {Vandenbussche}}]{Law2023}
{Law}, D.~R., {E. Morrison}, J., {Argyriou}, I., {et~al.} 2023, \aj, 166, 45, \dodoi{10.3847/1538-3881/acdddc}

\bibitem[{{Manara} {et~al.}(2023){Manara}, {Ansdell}, {Rosotti}, {Hughes}, {Armitage}, {Lodato}, \& {Williams}}]{Manara2023}
{Manara}, C.~F., {Ansdell}, M., {Rosotti}, G.~P., {et~al.} 2023, in Astronomical Society of the Pacific Conference Series, Vol. 534, Protostars and Planets VII, ed. S.~{Inutsuka}, Y.~{Aikawa}, T.~{Muto}, K.~{Tomida}, \& M.~{Tamura}, 539, \dodoi{10.48550/arXiv.2203.09930}

\bibitem[{{Mann} {et~al.}(2014){Mann}, {Di Francesco}, {Johnstone}, {Andrews}, {Williams}, {Bally}, {Ricci}, {Hughes}, \& {Matthews}}]{Mann2014}
{Mann}, R.~K., {Di Francesco}, J., {Johnstone}, D., {et~al.} 2014, \apj, 784, 82, \dodoi{10.1088/0004-637X/784/1/82}

\bibitem[{{Marigo} {et~al.}(2017){Marigo}, {Girardi}, {Bressan}, {Rosenfield}, {Aringer}, {Chen}, {Dussin}, {Nanni}, {Pastorelli}, {Rodrigues}, {Trabucchi}, {Bladh}, {Dalcanton}, {Groenewegen}, {Montalb{\'a}n}, \& {Wood}}]{Marigo2017}
{Marigo}, P., {Girardi}, L., {Bressan}, A., {et~al.} 2017, \apj, 835, 77, \dodoi{10.3847/1538-4357/835/1/77}

\bibitem[{{Mauc{\'o}} {et~al.}(2023){Mauc{\'o}}, {Manara}, {Ansdell}, {Bettoni}, {Claes}, {Alcala}, {Miotello}, {Facchini}, {Haworth}, {Lodato}, \& {Williams}}]{Mauco2023}
{Mauc{\'o}}, K., {Manara}, C.~F., {Ansdell}, M., {et~al.} 2023, \aap, 679, A82, \dodoi{10.1051/0004-6361/202347627}

\bibitem[{{McCaughrean} \& {O'Dell}(1996)}]{Mccaughrean1996}
{McCaughrean}, M.~J., \& {O'Dell}, C.~R. 1996, \aj, 111, 1977, \dodoi{10.1086/117934}

\bibitem[{{Miret-Roig} {et~al.}(2022){Miret-Roig}, {Galli}, {Olivares}, {Bouy}, {Alves}, \& {Barrado}}]{MiretRoig_2022}
{Miret-Roig}, N., {Galli}, P.~A.~B., {Olivares}, J., {et~al.} 2022, \aap, 667, A163, \dodoi{10.1051/0004-6361/202244709}

\bibitem[{{Miroshnichenko} {et~al.}(2013){Miroshnichenko}, {Pasechnik}, {Manset}, {Carciofi}, {Rivinius}, {{\v{S}}tefl}, {Gvaramadze}, {Ribeiro}, {Fernando}, {Garrel}, {Knapen}, {Buil}, {Heathcote}, {Pollmann}, {Mauclaire}, {Thizy}, {Martin}, {Zharikov}, {Okazaki}, {Gandet}, {Eversberg}, \& {Reinecke}}]{Miroshnichenko2013}
{Miroshnichenko}, A.~S., {Pasechnik}, A.~V., {Manset}, N., {et~al.} 2013, \apj, 766, 119, \dodoi{10.1088/0004-637X/766/2/119}

\bibitem[{{Morales} {et~al.}(2026){Morales}, {Bajaj}, {Pascucci}, \& {Clark}}]{ifuanalyzer}
{Morales}, J. S.~C., {Bajaj}, N.~S., {Pascucci}, I., \& {Clark}, S. 2026, {IFU-Analyzer}, GitHub.
\newblock \url{https://github.com/ilariapascucci/IFU-Analyzer}

\bibitem[{{Negueruela} {et~al.}(2024){Negueruela}, {Sim{\'o}n-D{\'\i}az}, {de Burgos}, {Casasbuenas}, \& {Beck}}]{Negueruela2024}
{Negueruela}, I., {Sim{\'o}n-D{\'\i}az}, S., {de Burgos}, A., {Casasbuenas}, A., \& {Beck}, P.~G. 2024, \aap, 690, A176, \dodoi{10.1051/0004-6361/202449298}

\bibitem[{Newville {et~al.}(2025)Newville, Otten, Nelson, Stensitzki, Ingargiola, Allan, Fox, Carter, \& Rawlik}]{lmfit}
Newville, M., Otten, R., Nelson, A., {et~al.} 2025, LMFIT: Non-Linear Least-Squares Minimization and Curve-Fitting for Python, 1.3.4,  Zenodo, \dodoi{10.5281/zenodo.16175987}

\bibitem[{{O'Dell}(1998)}]{Odell1998}
{O'Dell}, C.~R. 1998, \aj, 115, 263, \dodoi{10.1086/300178}

\bibitem[{{O'Dell}(2001)}]{Odell2001}
---. 2001, \araa, 39, 99, \dodoi{10.1146/annurev.astro.39.1.99}

\bibitem[{{O'Dell} {et~al.}(1993){O'Dell}, {Wen}, \& {Hu}}]{Odell1993}
{O'Dell}, C.~R., {Wen}, Z., \& {Hu}, X. 1993, \apj, 410, 696, \dodoi{10.1086/172786}

\bibitem[{{Otter} {et~al.}(2021){Otter}, {Ginsburg}, {Ballering}, {Bally}, {Eisner}, {Goddi}, {Plambeck}, \& {Wright}}]{Otter2021}
{Otter}, J., {Ginsburg}, A., {Ballering}, N.~P., {et~al.} 2021, \apj, 923, 221, \dodoi{10.3847/1538-4357/ac29c2}

\bibitem[{{Pabst} {et~al.}(2021){Pabst}, {Hacar}, {Goicoechea}, {Teyssier}, {Bern{\'e}}, {Wolfire}, {Higgins}, {Chambers}, {Kabanovic}, {G{\"u}sten}, {Stutzki}, {Kramer}, \& {Tielens}}]{Pabst2021}
{Pabst}, C.~H.~M., {Hacar}, A., {Goicoechea}, J.~R., {et~al.} 2021, \aap, 651, A111, \dodoi{10.1051/0004-6361/202140804}

\bibitem[{{Pascucci} {et~al.}(2023{\natexlab{a}}){Pascucci}, {Cabrit}, {Edwards}, {Gorti}, {Gressel}, \& {Suzuki}}]{Pascucci2023}
{Pascucci}, I., {Cabrit}, S., {Edwards}, S., {et~al.} 2023{\natexlab{a}}, in Astronomical Society of the Pacific Conference Series, Vol. 534, Protostars and Planets VII, ed. S.~{Inutsuka}, Y.~{Aikawa}, T.~{Muto}, K.~{Tomida}, \& M.~{Tamura} (ASP), 567, \dodoi{10.48550/arXiv.2203.10068}

\bibitem[{{Pascucci} {et~al.}(2023{\natexlab{b}}){Pascucci}, {Booth}, {Carpenter}, {Fang}, {Gorti}, {Liu}, {Long}, {Pontoppidan}, \& {Xie}}]{Pascucci2023prop}
{Pascucci}, I., {Booth}, R.~A., {Carpenter}, J., {et~al.} 2023{\natexlab{b}}, {The volatile content and C/O ratio of old disks: constraints on young planet atmospheres}, JWST Proposal. Cycle 2, ID. \#2970

\bibitem[{Pedregosa {et~al.}(2011)Pedregosa, Varoquaux, Gramfort, Michel, Thirion, Grisel, Blondel, Prettenhofer, Weiss, Dubourg, Vanderplas, Passos, Cournapeau, Brucher, Perrot, \& Duchesnay}]{scikit-learn}
Pedregosa, F., Varoquaux, G., Gramfort, A., {et~al.} 2011, Journal of Machine Learning Research, 12, 2825

\bibitem[{{Peri} {et~al.}(2012){Peri}, {Benaglia}, {Brookes}, {Stevens}, \& {Isequilla}}]{Peri2012}
{Peri}, C.~S., {Benaglia}, P., {Brookes}, D.~P., {Stevens}, I.~R., \& {Isequilla}, N.~L. 2012, \aap, 538, A108, \dodoi{10.1051/0004-6361/201118116}

\bibitem[{{Ping} {et~al.}(2026){Ping}, {Anania}, {Pinilla}, \& {Vioque}}]{Ping2026}
{Ping}, J., {Anania}, R., {Pinilla}, P., \& {Vioque}, M. 2026, \aap, 709, A6, \dodoi{10.1051/0004-6361/202558200}

\bibitem[{Ratzenb\"{o}ck {et~al.}(2023{\natexlab{a}})Ratzenb\"{o}ck, Gro{\ss}schedl, M\"{o}ller, Alves, Bomze, \& Meingast}]{ratzenbock2023a}
Ratzenb\"{o}ck, S., Gro{\ss}schedl, J.~E., M\"{o}ller, T., {et~al.} 2023{\natexlab{a}}, Astronomy \& Astrophysics, 677, a59, \dodoi{10.1051/0004-6361/202243690}

\bibitem[{Ratzenb\"{o}ck {et~al.}(2023{\natexlab{b}})Ratzenb\"{o}ck, Gro{\ss}schedl, Alves, Miret-Roig, Bomze, Forbes, Goodman, Hacar, Lin, Meingast, M\"{o}ller, Piecka, Posch, Rottensteiner, Swiggum, \& Zucker}]{ratzenbock2023b}
Ratzenb\"{o}ck, S., Gro{\ss}schedl, J.~E., Alves, J.~a., {et~al.} 2023{\natexlab{b}}, Astronomy \& Astrophysics, 678, a71, \dodoi{10.1051/0004-6361/202346901}

\bibitem[{{Raul} {et~al.}(2026){Raul}, {Zhang}, {Waggoner}, {Xie}, {Tallon}, {Banzatti}, {Salyk}, {Pontoppidan}, {Pascucci}, {Arulanantham}, {Vioque}, {Empey}, {Manara}, {Blake}, {Pinilla}, {Long}, {Yao}, {Kanwar}, {Bajaj}, {Jos{\'e} Colmenares}, {Kaeufer}, {Tabone}, {Bergin}, {Cieza}, {Narang}, {Miley}, {Krijt}, \& {Rosotti}}]{Raul2026}
{Raul}, E., {Zhang}, K., {Waggoner}, A., {et~al.} 2026, arXiv e-prints, arXiv:2606.27476, \dodoi{10.48550/arXiv.2606.27476}

\bibitem[{{Rilinger} \& {Espaillat}(2021)}]{Rilinger2021}
{Rilinger}, A.~M., \& {Espaillat}, C.~C. 2021, \apj, 921, 182, \dodoi{10.3847/1538-4357/ac09e5}

\bibitem[{{Robberto} {et~al.}(2013){Robberto}, {Soderblom}, {Bergeron}, {Kozhurina-Platais}, {Makidon}, {McCullough}, {McMaster}, {Panagia}, {Reid}, {Levay}, {Frattare}, {Da Rio}, {Andersen}, {O'Dell}, {Stassun}, {Simon}, {Feigelson}, {Stauffer}, {Meyer}, {Reggiani}, {Krist}, {Manara}, {Romaniello}, {Hillenbrand}, {Ricci}, {Palla}, {Najita}, {Ananna}, {Scandariato}, \& {Smith}}]{Robberto2013}
{Robberto}, M., {Soderblom}, D.~R., {Bergeron}, E., {et~al.} 2013, \apjs, 207, 10, \dodoi{10.1088/0067-0049/207/1/10}

\bibitem[{{Rogers} {et~al.}(2025){Rogers}, {de Marchi}, \& {Brandl}}]{Rogers2025}
{Rogers}, C., {de Marchi}, G., \& {Brandl}, B. 2025, \aap, 698, A172, \dodoi{10.1051/0004-6361/202453358}

\bibitem[{{Salyk} {et~al.}(2021){Salyk}, {Pontoppidan}, {Andrews}, {Banzatti}, {Bergin}, {Blake}, {Bosman}, {Carr}, {Krijt}, {Najita}, \& et~al.}]{Salyk2021}
{Salyk}, C., {Pontoppidan}, K.~M., {Andrews}, S.~M., {et~al.} 2021, {A DSHARP-MIRI Treasury survey of Chemistry in Planet-forming Regions}, JWST Proposal. Cycle 1, ID. \#1584

\bibitem[{{Scally} {et~al.}(2005){Scally}, {Clarke}, \& {McCaughrean}}]{Scally2005}
{Scally}, A., {Clarke}, C., \& {McCaughrean}, M.~J. 2005, \mnras, 358, 742, \dodoi{10.1111/j.1365-2966.2004.08617.x}

\bibitem[{{Sellek} {et~al.}(2020){Sellek}, {Booth}, \& {Clarke}}]{Sellek2020}
{Sellek}, A.~D., {Booth}, R.~A., \& {Clarke}, C.~J. 2020, \mnras, 492, 1279, \dodoi{10.1093/mnras/stz3528}

\bibitem[{{Smith} {et~al.}(2005){Smith}, {Bally}, {Shuping}, {Morris}, \& {Kassis}}]{Smith2005}
{Smith}, N., {Bally}, J., {Shuping}, R.~Y., {Morris}, M., \& {Kassis}, M. 2005, \aj, 130, 1763, \dodoi{10.1086/432912}

\bibitem[{{The pandas development Team}(2025)}]{pandas}
{The pandas development Team}. 2025, {pandas-dev/pandas: Pandas}, v2.3.1,  Zenodo, \dodoi{10.5281/zenodo.3509134}

\bibitem[{{Trapman} {et~al.}(2025){Trapman}, {Vioque}, {Kurtovic}, {Zhang}, {Rosotti}, {Pinilla}, {Carpenter}, {Cieza}, {Pascucci}, {Anania}, \& et~al.}]{Trapman2025}
{Trapman}, L., {Vioque}, M., {Kurtovic}, N.~T., {et~al.} 2025, \apj, 989, 10, \dodoi{10.3847/1538-4357/adc7af}

\bibitem[{{van der Plas} {et~al.}(2016){van der Plas}, {M{\'e}nard}, {Ward-Duong}, {Bulger}, {Harvey}, {Pinte}, {Patience}, {Hales}, \& {Casassus}}]{vanderplas2016}
{van der Plas}, G., {M{\'e}nard}, F., {Ward-Duong}, K., {et~al.} 2016, \apj, 819, 102, \dodoi{10.3847/0004-637X/819/2/102}

\bibitem[{{van Terwisga} \& {Hacar}(2023)}]{vanterwisga2023}
{van Terwisga}, S.~E., \& {Hacar}, A. 2023, \aap, 673, L2, \dodoi{10.1051/0004-6361/202346135}

\bibitem[{{van Terwisga} {et~al.}(2020){van Terwisga}, {van Dishoeck}, {Mann}, {Di Francesco}, {van der Marel}, {Meyer}, {Andrews}, {Carpenter}, {Eisner}, {Manara}, \& {Williams}}]{vanterwisga2020}
{van Terwisga}, S.~E., {van Dishoeck}, E.~F., {Mann}, R.~K., {et~al.} 2020, \aap, 640, A27, \dodoi{10.1051/0004-6361/201937403}

\bibitem[{Virtanen {et~al.}(2020)Virtanen, Gommers, Oliphant, Haberland, Reddy, Cournapeau, Burovski, Peterson, Weckesser, Bright, {van der Walt}, Brett, Wilson, Millman, Mayorov, Nelson, Jones, Kern, Larson, Carey, Polat, Feng, Moore, {VanderPlas}, Laxalde, Perktold, Cimrman, Henriksen, Quintero, Harris, Archibald, Ribeiro, Pedregosa, {van Mulbregt}, \& {SciPy 1.0 Contributors}}]{2020SciPy-NMeth}
Virtanen, P., Gommers, R., Oliphant, T.~E., {et~al.} 2020, Nature Methods, 17, 261, \dodoi{10.1038/s41592-019-0686-2}

\bibitem[{{Winter} {et~al.}(2019){Winter}, {Clarke}, {Rosotti}, {Hacar}, \& {Alexander}}]{Winter2019}
{Winter}, A.~J., {Clarke}, C.~J., {Rosotti}, G.~P., {Hacar}, A., \& {Alexander}, R. 2019, \mnras, 490, 5478, \dodoi{10.1093/mnras/stz2545}

\bibitem[{{Winter} \& {Haworth}(2022)}]{Winter2022}
{Winter}, A.~J., \& {Haworth}, T.~J. 2022, European Physical Journal Plus, 137, 1132, \dodoi{10.1140/epjp/s13360-022-03314-1}

\bibitem[{{Xie} {et~al.}(2026){Xie}, {Pascucci}, {Long}, {Gorti}, {Banzatti}, {Booth}, {Pontoppidan}, {Molyarova}, {Carpenter}, {Fang}, {Liu}, {Raul}, {Zhang}, {Ertel}, {Stone}, {Empey}, {Manara}, {Pinilla}, {Salyk}, {Tabone}, {Vioque}, {Cieza}, {Rosotti}, {Miley}, {Blake}, \& {Waggoner}}]{Xie2026}
{Xie}, C., {Pascucci}, I., {Long}, F., {et~al.} 2026, arXiv e-prints, arXiv:2606.27477, \dodoi{10.48550/arXiv.2606.27477}

\bibitem[{{Zallio} {et~al.}(2026{\natexlab{a}}){Zallio}, {Rosotti}, {Vioque}, {Miotello}, {Andrews}, {Manara}, {Carpenter}, {Empey}, {Kurtovic}, {Law}, \& et~al.}]{Zallio2026a}
{Zallio}, L., {Rosotti}, G.~P., {Vioque}, M., {et~al.} 2026{\natexlab{a}}, \aap, 705, A49, \dodoi{10.1051/0004-6361/202557366}

\bibitem[{{Zallio} {et~al.}(2026{\natexlab{b}}){Zallio}, {Vioque}, {Andrews}, {Empey}, {Rosotti}, {Miotello}, {Manara}, {Carpenter}, {Deng}, {Kurtovic}, {Law}, {Longarini}, {Paneque-Carre{\~n}o}, {Teague}, {Villenave}, {Yen}, \& {Zagaria}}]{Zallio2026b}
{Zallio}, L., {Vioque}, M., {Andrews}, S.~M., {et~al.} 2026{\natexlab{b}}, \aap, 708, L1, \dodoi{10.1051/0004-6361/202558147}

\bibitem[{{Zhang} {et~al.}(2023){Zhang}, {Anderson}, {Beatty}, {Bergin}, {Blake}, {Carpenter}, {Cieza}, {Hogerheijde}, {Miley}, {Pascucci}, \& et~al.}]{Zhang2023}
{Zhang}, K., {Anderson}, D., {Beatty}, T.~G., {et~al.} 2023, {Building on ALMA: a JWST legacy survey of the chemical evolution of planet-forming disks}, JWST Proposal. Cycle 2, ID. \#3034

\end{thebibliography}
\bibliographystyle{aasjournal}

\end{document}